\documentclass{iopjournal}

\usepackage{amsmath,amssymb}
\usepackage{graphicx}
\usepackage{bm}
\usepackage{amsthm}
\usepackage[dvipsnames]{xcolor}
\hypersetup{
    colorlinks,
    linkcolor={red!50!black},
    citecolor={blue!50!black},
    urlcolor={blue!80!black}
}

\makeatletter
\newcommand*{\rom}[1]{\expandafter\@slowromancap\romannumeral #1@}
\makeatother

\usepackage{hyperref}

\begin{document}

\title{Higher–curvature corrections and the endpoint of black hole evaporation in gravitational effective field theory}

\author{Lorens F. Niehof$^{1,2,*}$\orcid{0009-0005-3143-5730}, Sjors Heefer$^1$\orcid{0000-0002-2057-6301}, Andrea Fuster$^1$\orcid{0000-0001-6095-6838} and Federico Toschi$^{2}$\orcid{0000-0001-5935-2332}}

\affil{$^1$Department of Mathematics and Computer Science \\
 Eindhoven University of Technology, Eindhoven 5600MB, The Netherlands}

\affil{$^2$Department of Applied Physics and Science Education \\
 Eindhoven University of Technology, Eindhoven 5612AP, The Netherlands}

\affil{$^*$Author to whom any correspondence should be addressed.}

\email{l.f.niehof@student.tue.nl}


%Black Hole Evaporation, Hawking Radiation, Black Hole Thermodynamics, Effective Field Theory of Gravity, Higher-Curvature Corrections, Black Hole Remnants

\begin{abstract}

{\small The endpoint of black hole evaporation remains uncertain once the semiclassical description approaches the Planck scale. In this work we study late–stage evaporation within four–dimensional gravitational effective field theory. We consider the leading local correction to the Schwarzschild solution arising from a cubic curvature operator, and use the corrected geometry to analyze the resulting evaporation dynamics and associated thermodynamic properties.

We show that the cubic correction induces a parametric slow–down of the evaporation rate at small masses, which within the truncated theory can appear as a freeze–out at a finite mass scale. We demonstrate that this behavior is not an independent physical prediction, but instead occurs precisely when the dimensionless expansion parameter of the effective theory becomes of order unity. The corresponding mass scale coincides parametrically with the onset of Planckian curvature at the horizon, establishing that the evaporation dynamics provide a direct diagnostic of the breakdown of the effective field theory.

A scaling analysis of higher–order curvature operators shows that once the cubic term becomes comparable to the Einstein–Hilbert contribution, generic higher–order terms are no longer parametrically suppressed. The apparent remnant–like behavior therefore arises at the boundary of validity of the effective description rather than within a controlled perturbative regime. These results demonstrate that late–stage evaporation encodes the limits of gravitational effective field theory, providing a dynamical criterion for its breakdown.}

\end{abstract}

\section{Introduction}\label{sec:introduction}

\noindent In his groundbreaking 1974 paper, Hawking predicted that black holes should radiate as if they were black bodies \cite{Hawking1974Explosions}. As a consequence, an isolated black hole loses energy and thus mass through Hawking radiation. In the absence of accretion, this process leads to complete evaporation in finite time within the semiclassical description. Hawking's predictions initiated the modern understanding of black hole thermodynamics and have had profound implications for quantum gravity.

However, the ultimate endpoint of black hole evaporation remains uncertain. More recently, it has been proposed that full evaporation may not be the final stage of a black hole \cite{MacGibbon1987Close, Bowick1988Axionic, Barrow1992Cosmology, Banks1992Horned, Whitt1988Spherically, Alexeyev2002String, Giddings1992Massive, Aharonov1987Unitarity, Banks1993Information, Callan1989String, Myers1988Lovelock, Chen2015Information}. This possibility is often motivated by the black hole information paradox \cite{Aharonov1987Unitarity, Banks1993Information, Callan1989String, Myers1988Lovelock, Chen2015Information}: after a black hole is formed and then evaporates fully through Hawking radiation, its final state can be described only through the mass, electric charge, and angular momentum of the initial state. This suggests that distinct initial states could evolve into the same final state, apparently violating unitarity and suggesting that information about the initial state is lost.

One frequently discussed way to address this tension is to postulate the existence of a black hole remnant, typically of Planck-scale mass and size\footnote{That is, a black hole with mass of order $M_p = \sqrt{\frac{\hbar c}{G}} \approx 2.176 \cdot 10^{-8} \text{ kg}$ and radius of order $\ell_p = \sqrt{\frac{\hbar G}{c^3}} \approx 1.616 \cdot 10^{-35} \text{ m}$.}. Such remnants, often referred to as Planck relics, have been extensively discussed in the literature\footnote{Some works further speculate that Planck relics could contribute to the dark matter abundance. This possibility will not be explored here; see \cite{Barrau2019DarkMatter} for a review.} \cite{MacGibbon1987Close, Barrow1992Cosmology, Whitt1988Spherically, Alexeyev2002String, Giddings1992Massive, Aharonov1987Unitarity, Banks1993Information, Chen2015Information}. In many cases, remnants are introduced phenomenologically, without derivation from a controlled gravitational framework. An example is the work of Barrow et al. \cite{Barrow1992Cosmology}, where a modified Hawking temperature is postulated and shown to lead to evaporation freeze-out. While that construction is explicitly phenomenological, it demonstrates how modest deviations from semiclassical thermodynamics can qualitatively alter the endpoint of evaporation.

A complementary and systematically controlled approach is provided by gravitational effective field theory (EFT). At energy scales well below the Planck scale, gravity can be described by an effective action organized as a derivative expansion in local curvature invariants,
\begin{equation}
    S = \int d^4 x \sqrt{-g} \left[ \frac{M_p^2}{16\pi} R + c_1 R^2 + c_2 R_{\mu\nu}R^{\mu\nu} + c_3 R_{\mu \nu \rho \sigma} R^{\mu \nu \rho \sigma}+ \frac{c_6}{M_p^2} \mathcal{R}^3 + \frac{c_8}{M_p^4} \mathcal{R}^4 + \cdots \right],
\end{equation}
where $g$ is the determinant of the metric tensor $g_{\mu \nu}$, and $\mathcal{R}^n$ schematically denotes contractions of $n$ curvature tensors. The coefficients $c_i$ are dimensionless Wilson coefficients encoding the effects of ultraviolet physics \cite{Donoghue1994Leading}.

Gravitational effective field theory provides the unique low-energy description of quantum gravity consistent with general covariance, independent of the details of the ultraviolet completion \cite{Donoghue1994Leading,Burgess2004Everyday}. In this framework, all operators compatible with the symmetries of the theory appear in the effective action, organized according to their mass dimension. The unknown ultraviolet completion enters only through the Wilson coefficients $c_i$, while the structure of the expansion itself is universal. Consequently, although the effective theory is not expected to remain quantitatively reliable once Planckian curvatures are reached, it provides a systematically improvable and model-independent description of gravitational dynamics throughout the regime in which the derivative expansion remains perturbative.

The expansion is controlled by the dimensionless ratio $R/M_p^2$, so that each higher-order term is suppressed by additional powers of curvature relative to the Planck scale. For black hole spacetimes, the curvature near the horizon scales as $R \sim 1/r_H^2 \sim M_p^4/M^2$, implying
\begin{equation}
    \frac{R}{M_p^2} \sim \left(\frac{M_p}{M}\right)^2.
\end{equation}
The derivative expansion therefore becomes an expansion in powers of $(M_p/M)^2$, which grows as the black hole mass approaches the Planck scale.

In this work, we revisit the question of late-stage evaporation within the EFT framework. Rather than postulating modifications of black hole thermodynamics, we derive leading corrections from a local third-order curvature expansion of the gravitational action, building on the cubic-curvature corrections to the Schwarzschild solution obtained in \cite{Calmet2021Entropy}. Within this framework, the leading correction is controlled by the dimensionless parameter
\begin{equation}
    \epsilon \sim c_6 \left(\frac{M_p}{M}\right)^4,
\end{equation}
which corresponds to the ratio of the cubic curvature term to the Einstein--Hilbert contribution.

While it is generally expected that the effective field theory breaks down when curvature reaches the Planck scale, this expectation is usually formulated at a kinematical level. Here, we instead investigate how this breakdown is reflected dynamically in the evaporation process. We show that the evaporation dynamics itself becomes sensitive to the loss of perturbative control, and that the slow–down or apparent freeze–out of evaporation arises precisely when the expansion parameter $\epsilon$ becomes of order unity.

This provides a direct dynamical signature of the breakdown of the effective theory. The corresponding mass scale coincides parametrically with the onset of Planckian curvature at the horizon, implying that the remnant–like behavior of the truncated description does not constitute a prediction within its regime of validity, but rather signals the breakdown of the derivative expansion.

The structure of this paper is as follows. In Section~\ref{sec:literature} we provide a focused review of higher–curvature black hole thermodynamics and evaporation in the literature, clarifying the conceptual context and positioning the present work relative to previous studies. In Section~\ref{sec:EFT} we introduce the effective field theory framework and summarize how the third–order curvature term modifies the Schwarzschild geometry. In Section~\ref{sec:evaporation} we identify the emergence of evaporation slow–down and freeze–out behavior within the truncated EFT and derive the associated characteristic mass scale. Section~\ref{sec:thermo} analyzes the corresponding thermodynamic properties, including the time-dependent behavior of the Hawking temperature, and the sign-dependent structure of the evaporation time and the heat capacity. In Section~\ref{sec:validity} we examine the regime of validity of the derivative expansion by evaluating curvature invariants at the critical mass and comparing the cubic correction to quartic and higher–order operators in the action. Then, in Section~\ref{sec:extensions} we discuss extensions by including the role of greybody factors, and discussing qualitative considerations for charged and rotating black holes. Finally, in Section~\ref{sec:comparison} we directly compare the EFT results of this work with other approaches to quantum gravity. Throughout, we carefully distinguish results that are parametrically controlled within the EFT from features that arise at the boundary of its applicability. In this sense, the late stages of black hole evaporation act as a probe of the limits of the gravitational effective field theory.

\section{Higher–Curvature Black Hole Thermodynamics and Evaporation: Literature Overview}\label{sec:literature}
The study of black hole thermodynamics in theories containing higher–curvature corrections has a long history, motivated both by quantum gravity and by EFT considerations. Since local curvature invariants necessarily arise in the low–energy expansion of any ultraviolet completion of gravity \cite{Donoghue1994Leading, Burgess2004Everyday}, it is natural to ask how such terms modify black hole solutions, their thermodynamic properties, and ultimately their evaporation dynamics.

\subsection{Quadratic curvature gravity and Lovelock theories}
Early investigations of higher–curvature modifications of gravity focused primarily on quadratic curvature terms, including invariants such as $R^2$, $R_{\mu\nu}R^{\mu\nu}$, and $R_{\mu\nu\rho\sigma}R^{\mu\nu\rho\sigma}$. These corrections arise naturally in attempts to quantize gravity and were first studied systematically in the context of renormalizable higher–derivative gravity by Stelle \cite{Stelle1977Renormalization}. 

In higher dimensions, particular combinations of quadratic invariants (notably the Gauss–Bonnet term $R^2 - 4 R^{\mu \nu} R_{\mu \nu} + R^{\mu \nu \rho \sigma}R_{\mu \nu \rho \sigma}$) lead to the broader class of Lovelock theories, which admit exact black hole solutions with modified thermodynamic properties \cite{Wheeler1986Lovelock, MyersSimon1988GB}. In these settings, higher–curvature corrections can significantly alter the temperature, entropy, and heat capacity of black holes, and can introduce additional solution branches and modified stability behavior \cite{Cai2004GBThermo, PadmanabhanKothawala2013}. 

In four spacetime dimensions, however, quadratic curvature terms do not modify the Schwarzschild solution in vacuum at leading order. This follows from the fact that such corrections either vanish on Ricci–flat backgrounds or can be removed by field redefinitions at the level of the effective action. Consequently, the first nontrivial local corrections to Schwarzschild geometry within a four–dimensional effective field theory arise only at cubic order in curvature \cite{Calmet2021Entropy, LuPerkinsPopeStelle2015, Calmet2018Quantum, Calmet2018VanishingCurvature}. 

In quadratic and Lovelock theories, higher–curvature terms can also generate extremal configurations with vanishing Hawking temperature and qualitatively modified thermodynamic evolution \cite{CveticNojiriOdintsov2002}. However, these analyses are typically performed either in higher dimensions or within exact modified gravity theories treated nonperturbatively, rather than within a controlled four–dimensional effective field theory expansion. 

For this reason, cubic curvature operators provide the leading setting in which one can investigate genuinely new thermodynamic effects of higher–curvature gravity on four–dimensional black holes within a systematic perturbative EFT framework.

\subsection{Cubic curvature gravity and Einsteinian cubic gravity}
Renewed interest in cubic curvature invariants followed the construction of four–dimensional theories admitting nontrivial black hole solutions while retaining second–order linearized equations, most notably Einsteinian cubic gravity (ECG) \cite{BuenoCano2016, Hennigar2017}. In these works, static and asymptotically AdS black hole solutions were obtained and their thermodynamic properties analyzed in detail. Notably, cubic corrections were shown to modify the Hawking temperature, entropy (via the Wald formula), and heat capacity, sometimes producing bounded temperature behavior.

Subsequent work extended these analyses to rotating solutions and to detailed studies of thermodynamic stability \cite{Hennigar2017, BuenoCano2017}. Importantly, in some regimes small black holes were found to exhibit a maximal temperature followed by cooling as the horizon radius decreases \cite{BuenoCano2017}, suggesting the possibility of long–lived configurations within those specific theories.

However, most of this literature treats cubic gravity as a complete modified theory rather than as the leading term in an EFT expansion. In particular, the cubic couplings are often taken to be finite and not parametrically suppressed relative to the Einstein–Hilbert term. As emphasized in later analyses \cite{DeFelice2023, Gonzalez2023}, such treatments can lead to ghostlike excitations or instabilities unless the cubic operators are interpreted perturbatively within EFT.

\subsection{Thermodynamic corrections in gravitational effective field theory}
Within the EFT of gravity, higher–curvature operators arise with Wilson coefficients suppressed by appropriate powers of the ultraviolet scale \cite{Donoghue1994Leading, Burgess2004Everyday}. In this framework, thermodynamic quantities can be computed perturbatively around classical solutions. Calmet et al. derived the cubic–corrected Schwarzschild metric to first order in the EFT expansion parameter $\epsilon \sim c_6 (M_p / M)^4$ and computed the associated corrections to the Hawking temperature and entropy \cite{Calmet2021Entropy}. These results provide a controlled starting point for analyzing how cubic operators modify semiclassical black hole physics in asymptotically flat spacetime.

A key structural feature of EFT is that the expansion is organized in powers of curvature. Once the dimensionless parameter $\epsilon$ controlling the derivative expansion becomes of order unity, the truncation ceases to be hierarchical and higher–order operators contribute at the same parametric order. While this general expectation is well understood, the detailed interplay between cubic corrections, evaporation dynamics, and the breakdown of the derivative expansion has not been systematically analyzed in the context of asymptotically flat Hawking evaporation. In particular, it has not been shown how the evaporation dynamics itself reflects this breakdown within a controlled EFT framework.

\subsection{Higher–curvature effects and evaporation endpoints}
The possibility that higher–curvature corrections might alter the endpoint of evaporation has been discussed in several frameworks. In asymptotic safety, renormalization–group improved black hole metrics can yield a maximal temperature and vanishing temperature at finite mass \cite{Bonanno2000,Bonanno2006}. In Gauss–Bonnet and Lovelock theories, extremal or minimal–mass configurations can arise depending on the sign of the coupling \cite{Cai2002}. In generalized uncertainty principle and nonlocal gravity models, modified dispersion relations or nonlocal form factors can similarly suppress late–time evaporation \cite{Adler2001,Scardigli1999,Modesto2010,Biswas2012}.

Despite these developments, two important issues remain open. First, most analyses of cubic or higher–curvature thermodynamics focus on exact modified gravity theories. Second, while modified temperature behavior has been observed in several models, the relation between evaporation slow–down and the regime of validity of the gravitational effective field theory expansion has not been systematically clarified. These considerations motivate a closer examination of higher–curvature corrections within a controlled EFT framework.

\subsection{Position and novelty of the present work}
The present work differs from previous studies of higher–curvature black hole thermodynamics in several important respects.

(i) We work strictly within a four–dimensional local gravitational EFT, treating the cubic curvature operator perturbatively as the leading correction to the quadratic action following \cite{Calmet2021Entropy}. Unlike many modified–gravity approaches, the higher–curvature term is not interpreted as defining a complete theory, but rather as the leading operator in a derivative expansion valid below the Planck scale.

(ii) Within this framework we analyze not only static thermodynamic quantities but also the evaporation dynamics of asymptotically flat black holes. Using the corrected Hawking temperature and horizon area, we derive the modified mass–loss law and study how higher–curvature corrections affect the late stages of evaporation.

(iii) We identify a direct link between evaporation dynamics and EFT validity by showing that the characteristic mass scale at which evaporation slow–down or freeze–out appears coincides parametrically with the breakdown of the derivative expansion. This establishes the evaporation dynamics itself as a diagnostic of EFT breakdown.

(iv) A parametric scaling analysis of higher–order curvature operators shows that once the cubic correction becomes comparable to the Einstein–Hilbert contribution, the derivative expansion ceases to be hierarchical and generic higher–order operators contribute at the same order. This demonstrates that the apparent freeze–out arises precisely at the boundary of EFT validity and cannot be consistently interpreted within the truncated theory.

\vspace{4mm}

In this way, the present work does not simply reproduce modified thermodynamic behavior in higher–curvature gravity. Instead, it clarifies how the structure of the gravitational effective field theory itself constrains the interpretation of late–stage evaporation dynamics and apparent remnant formation.

\section{Effective Field Theory}\label{sec:EFT}
Motivated by the considerations outlined in the previous section, we now formulate the problem within the framework of gravitational effective field theory (EFT). In this approach, higher–curvature operators encode the leading quantum corrections to the classical Einstein–Hilbert action. The low–energy dynamics are organized as a derivative expansion in powers of curvature, with higher–order terms suppressed by the appropriate ultraviolet scale. For reviews of gravity as an effective field theory, see e.g. \cite{Burgess2004Everyday, Donoghue2012Effective, Saraswat2017Weak, Ruhdorfer2020Effective}.

As discussed in Section \ref{sec:literature}, no corrections to the Schwarzschild metric arise at quadratic order in curvature in vacuum in four dimensions \cite{Calmet2021Entropy, LuPerkinsPopeStelle2015, Calmet2018Quantum, Calmet2018VanishingCurvature}. The leading local modifications therefore appear at cubic order in curvature. These operators provide the first nontrivial corrections to classical Schwarzschild geometry within the four–dimensional EFT expansion.

It is important to stress that the EFT treatment is intrinsically perturbative, and its validity requires that the curvature invariants remain small compared to the cutoff scale. As a result, the expressions derived below should be interpreted as asymptotic expansions whose breakdown signals the onset of genuinely nonperturbative quantum-gravitational effects. In what follows, we therefore treat all derived quantities as asymptotic indicators of qualitative behavior, rather than as controlled predictions arbitrarily close to the Planck scale.

We now summarize the effective field theory action relevant for our analysis. We first present the local third-order terms that contribute in vacuum, followed by a brief discussion of nonlocal corrections. We then outline how the local terms modify the Schwarzschild metric, which forms the basis for the subsequent thermodynamic analysis.

\subsection{The Effective Action}

Following \cite{Calmet2021Entropy}, we consider the local gravitational effective action in vacuum:
\begin{align}\label{eq:EFTaction}
    S_{EFT,L} = &\int d^4 x \sqrt{-g} \left(\frac{M_p^2}{16\pi}R + c_1 R^2 + c_2 R_{\mu \nu} R^{\mu \nu} + c_3 R_{\mu \nu \rho \sigma} R^{\mu \nu \rho \sigma} + \frac{c_6}{M_p^2} R^{\mu \nu}{}_{\alpha \sigma} R^{\alpha \sigma}{}_{\delta \gamma}R^{\delta \gamma}{}_{\mu \nu}\right).
\end{align}
The Einstein-Hilbert term $\frac{M_p^2}{16\pi}R$ yields the classical field equations, while the quadratic terms $c_1 R^2, ~c_2 R_{\mu \nu} R^{\mu \nu},~ c_3 R_{\mu \nu \rho \sigma} R^{\mu \nu \rho \sigma}$ constitute the most general local curvature-squared corrections. The cubic term $\frac{c_6}{M_p^2} R^{\mu \nu}{}_{\alpha \sigma} R^{\alpha \sigma}{}_{\delta \gamma}R^{\delta \gamma}{}_{\mu \nu}$ represents the leading higher-curvature operator that contributes nontrivially in vacuum at this order.

The sign and magnitude of the Wilson coefficient $c_6$ depend on the ultraviolet completion of gravity. In particular, string-inspired models generate higher-curvature operators whose coefficients can take either sign \cite{Goroff1986Ultraviolet, Calmet2017QuantumCorrections}. No general principle is currently known that fixes the sign of $c_6$, so both possibilities must be considered.

This $c_6$ term is the only independent cubic-in-curvature invariant that contributes in vacuum, up to total derivatives and terms vanishing by the Bianchi identities \cite{Goroff1986Ultraviolet}. While quartic and higher-order operators can in principle be included, we restrict attention to cubic order, which already suffices to illustrate how higher-curvature effects can qualitatively modify black hole thermodynamics.

The nonlocal part of the effective action is given by
\begin{align}\label{eq:EFTnonlocal}
    S_{EFT,NL} = - &\int d^4 x \sqrt{-g} \left(C_1 R \ln{\left(\frac{\Box}{\mu^2}\right)} R + C_2 R_{\mu \nu}\ln{\left(\frac{\Box}{\mu^2}\right)}R^{\mu \nu} \right.  \\
    &\left.+ C_3 R_{\mu \nu \alpha \beta} \ln{\left(\frac{\Box}{\mu^2}\right)}R^{\mu \nu \alpha \beta} + C_4 R^{\mu \nu}{}_{\alpha \sigma}\log{\Box}R^{\alpha \sigma}{}_{\delta \gamma}R^{\delta \gamma}{}_{\mu \nu}\right), \nonumber
\end{align}
where the $C_i$ are coupling constants, $\Box = g^{\mu \nu} \nabla_{\mu} \nabla_{\nu}$, and $\mu$ denotes the renormalization scale.

Nonlocal terms encode infrared quantum effects arising from loops of massless fields and ensure renormalization group invariance of the EFT \cite{Donoghue1994Leading, Burgess2004Everyday}. Because our focus is on near-horizon, high-curvature modifications of black hole geometry, we neglect nonlocal terms in the present analysis, following \cite{Calmet2021Entropy}. This restriction should be viewed as a simplifying assumption rather than a statement of completeness.

\subsection{Metric Corrections}\label{sec:metriccorrections}

We now summarize how the cubic curvature term modifies the Schwarzschild geometry. The metric derived in \cite{Calmet2021Entropy} is obtained perturbatively to first order in the expansion parameter
\begin{equation}
    \epsilon \sim c_6 \left(\frac{M_p}{M}\right)^4.
\end{equation}
and takes the form
\begin{align}\label{eq:SchwarzschildAdapted}
    ds^2 = -f(r)dt^2 + \frac{1}{g(r)}dr^2 + r^2 d\Omega^2,
\end{align}
with
\begin{align}
    f(r) &= 1 - 2 \left(\frac{\ell_p}{r}\right)\left(\frac{M}{M_p}\right) + 640 \pi c_6 \left(\frac{\ell_p}{r}\right)^7\left(\frac{M}{M_p}\right)^3, \\
    g(r) &= 1 - 2 \left(\frac{\ell_p}{r}\right)\left(\frac{M}{M_p}\right) + 128 \pi c_6 \left(\frac{\ell_p}{r}\right)^6 \left(\frac{M}{M_p}\right)^2\left(27-49\left(\frac{\ell_p}{r}\right)\left(\frac{M}{M_p}\right)\right).
\end{align}

The location of the Killing horizon is defined by $f(r_H)=0$. While this equation cannot be solved analytically, one can determine $r_H$ perturbatively to first order in the expansion parameter $\epsilon \sim c_6 (M_p/M)^4$:
\begin{align}\label{eq:horradcor}
    r_H &= 2 \ell_p \left(\frac{M}{M_p}\right) \left(1-5\pi c_6 \left(\frac{M_p}{M}\right)^4 \right)
    + \mathcal{O}\left(c_6^2 \left(\frac{M_p}{M}\right)^8\right).
\end{align}

The corresponding horizon area becomes
\begin{align}
    A_H = 4\pi r_H^2
    = 16\pi \ell_p^2 \left(\frac{M}{M_p}\right)^2
    \left[1 - 10\pi c_6 \left(\frac{M_p}{M}\right)^4 \right]
    + \mathcal{O}\left(c_6^2 \left(\frac{M_p}{M}\right)^8\right).
\end{align}

These expressions are reliable only as long as the expansion in $c_6 (M_p/M)^4$ remains perturbative. While the corrections are entirely negligible for astrophysical black holes, they become increasingly important as $M$ approaches the Planck scale, where the EFT expansion itself is expected to break down. In the following sections, we use these results to explore the qualitative impact of higher-curvature corrections on black hole thermodynamics, while explicitly keeping track of the limitations of the perturbative treatment.

\subsection{Perturbative consistency of the corrected horizon}
Before proceeding to the thermodynamic analysis, it is useful to comment on the geometric nature of the corrected horizon. Since the metric \eqref{eq:SchwarzschildAdapted} is obtained as a perturbative solution of the effective field equations, it is only valid to first order in the cubic curvature coupling $c_6$. Consequently, all geometric quantities must be interpreted consistently within this perturbative expansion, while terms of order $\mathcal{O}\!\left(c_6^2(M_p/M)^8\right)$ lie beyond the accuracy of the effective description.

The location of the horizon is determined perturbatively by the largest root of
\begin{equation}
    f(r_H)=0,
\end{equation}
which yields the expression of Eq. \eqref{eq:horradcor}. Substituting this expression into the metric function $g(r)$ gives
\begin{equation}
    g(r_H) = 0 + \mathcal{O}\!\left(c_6^2\left(\frac{M_p}{M}\right)^8\right),
\end{equation}
such that $g(r_H)$ vanishes to the perturbative order retained throughout this work. The normal vector to the hypersurface $r=r_H$ is given by
\begin{equation}
    n_\mu=\partial_\mu(r-r_H),
\end{equation}
with norm
\begin{equation}
    n^\mu n_\mu = g^{rr} = g(r_H).
\end{equation}

Hence, within the accuracy of the effective field theory solution, the hypersurface $r=r_H$ remains null. Since the spacetime is static with Killing vector $\xi=\partial_t$, the Killing vector satisfies $\xi^{\mu} \xi_{\mu} = -f(r_H)=\mathcal{O}\left(c_6^2\left(\frac{M_p}{M}\right)^8\right)$ on the perturbatively corrected horizon. The standard expression for the surface gravity therefore remains applicable,
\begin{equation}
    \kappa = \frac{1}{2} \sqrt{f'(r_H)g'(r_H)},
\end{equation}
and consequently the Hawking temperature is given by
\begin{equation}
    T_H = \frac{\kappa}{2\pi} = \frac{\sqrt{f'(r_H)g'(r_H)}}{4\pi},
\end{equation}
consistent with the perturbative accuracy of the metric.

It is also instructive to verify that the metric retains its Lorentzian signature across the corrected horizon. Expanding the derivatives of the metric functions about $r=r_H$ yields
\begin{align}
    f'(r_H) &= \frac{M_p}{2\ell_p M} - \frac{25\pi c_6}{2\ell_p} \frac{M_p^5}{M^5} + \mathcal{O}\!\left(c_6^2\left(\frac{M_p}{M}\right)^9\right),\\
    g'(r_H) &= \frac{M_p}{2\ell_p M} + \frac{29\pi c_6}{2\ell_p} \frac{M_p^5}{M^5} + \mathcal{O}\!\left(c_6^2\left(\frac{M_p}{M}\right)^9\right).
\end{align}
The higher-curvature corrections are suppressed by the expansion parameter $c_6(M_p/M)^4$, so that within the perturbative regime of validity of the effective field theory both derivatives remain positive. Consequently, $f(r)$ and $g(r)$ change sign simultaneously across $r=r_H$, exactly as in the Schwarzschild solution, and the metric therefore preserves its Lorentzian signature to the perturbative order considered. Any possible separation between the zeros of $f(r)$ and $g(r)$ would necessarily arise only at $\mathcal{O}\!\left(c_6^2(M_p/M)^8\right)$, which lies beyond the accuracy of the present effective description.

Finally, we have explicitly verified that the corrected horizon is not associated with a curvature singularity. To first order in the cubic curvature correction, the Kretschmann scalar is
\begin{equation}
    K = R_{\mu\nu\rho\sigma}R^{\mu\nu\rho\sigma} = \frac{48\ell_p^2M^2}{r^6M_p^2} + \frac{18432\pi c_6\,\ell_p^7M^3 \left(25\ell_pM-12rM_p\right)} {r^{12}M_p^4} + \mathcal{O}\!\left(c_6^2\right).
\end{equation}

Evaluating this expression on the corrected horizon gives
\begin{equation}
    K(r_H) = \frac{3}{4\ell_p^4}\left(\frac{M_p}{M}\right)^4 \left[1+ 36\pi c_6 \left(\frac{M_p}{M}\right)^4\right] + \mathcal{O}\left(c_6^2\left(\frac{M_p}{M}\right)^{12}\right),
\end{equation}
which is manifestly finite. Therefore, to the perturbative order considered here, the corrected horizon does not coincide with a curvature singularity, and the only curvature singularity appearing within the perturbative regime described by the effective field theory remains at $r=0$, as in the Schwarzschild spacetime.

Finally, we comment on the possible existence of additional horizons. The perturbative solution considered here is constructed as a deformation of the Schwarzschild geometry and is valid only to first order in the cubic curvature coupling. Within this perturbative framework, the event horizon is obtained by continuously deforming the Schwarzschild horizon, yielding the corrected radius given in Eq.~\eqref{eq:horradcor}. The effective field theory therefore determines only the perturbative displacement of this horizon.

Although the truncated metric functions may formally admit additional roots when extrapolated into the high-curvature regime, such roots would occur only at radii where the higher-curvature expansion is no longer parametrically controlled and neglected operators become equally important. Their existence is therefore not physically meaningful within the truncated effective field theory, and they cannot be interpreted as genuine inner or Cauchy horizons. Consequently, the present solution does not provide a controlled setting in which phenomena such as mass inflation or Cauchy horizon instabilities can be analyzed. Establishing whether such structures arise would require knowledge of the complete nonperturbative geometry beyond the regime of validity of the cubic EFT expansion.

These results demonstrate that the perturbatively corrected solution preserves the essential geometric properties required for the thermodynamic analysis carried out in the remainder of this work. In particular, the corrected horizon remains a null hypersurface within the accuracy of the effective field theory expansion, the spacetime retains its Lorentzian character across the horizon, and no additional curvature singularities appear at the horizon itself.

\section{Identification of Evaporation Freeze-Out Behavior}\label{sec:evaporation}

In this section, we investigate how higher-curvature corrections within the truncated effective field theory modify the late-stage behavior of black hole evaporation. Rather than asserting the existence of physical remnants, we identify conditions under which the evaporation dynamics exhibit freeze-out or slow-down behavior within the perturbative framework.

Black hole remnants are often heuristically associated with a halting of evaporation, preventing further mass loss through Hawking radiation. Within the present truncated framework, such behavior can be identified in two distinct ways:
\begin{itemize}
    \item The corrected Hawking temperature tends to zero\footnote{For stationary black holes the Hawking temperature is determined by the surface gravity $\kappa$ at the event horizon through $T_H = \kappa/(2\pi)$ \cite{Hawking1975}. In the Schwarzschild case one has $\kappa = 1/(4GM)$, so that $T_H = 1/(8\pi GM)$. A vanishing Hawking temperature therefore corresponds to a vanishing surface gravity. In the perturbatively corrected solutions considered here, following \cite{Calmet2021Entropy}, the Hawking temperature continues to be obtained from the surface gravity of the corrected horizon geometry, so that the condition $T_H \to 0$ remains equivalent to $\kappa \to 0$ within the truncated effective field theory.}, suppressing thermal radiation.
    \item The mass-loss rate implied by the Stefan–Boltzmann law satisfies $\frac{dM}{dt}=0$ at a nonzero mass $M_{crit}$.
\end{itemize}

In standard semiclassical evaporation these criteria coincide, since the emitted power scales as $T_H^4$. Once effective field theory corrections are included and expanded perturbatively, however, this equivalence need not persist, and the two criteria may signal qualitatively different mechanisms for evaporation slow-down.

In the remainder of this section, we treat separately the cases $c_6 < 0$ and $c_6 > 0$. We show that, within the truncated EFT expansion, each case exhibits a form of evaporation freeze-out, while emphasizing that the precise endpoint lies beyond the regime of strict EFT control.

\subsection{Modified Hawking Temperature and Freeze-Out Behavior for \texorpdfstring{$c_6 < 0$}{}}

We begin with the case $c_6 < 0$, where the effective field theory correction drives the Hawking temperature toward zero at a finite mass within the perturbative expansion. The Hawking temperature derived in \cite{Calmet2021Entropy} to first order in $c_6 \left(M_p / M\right)^4$ is
\begin{align}\label{eq:hawktempcor}
    T_H = \frac{T_p}{8\pi} \left(\frac{M_p}{M}\right)&\left[1 + 2\pi c_6\left(\frac{M_p}{M}\right)^4 \right] + \mathcal{O}\left(c_6^2 \left(\frac{M_p}{M}\right)^8\right).
\end{align}

This expression is not postulated but follows from the perturbative quantum-corrected Schwarzschild solution of \cite{Calmet2021Entropy}. There, the surface-gravity definition of the temperature remains valid once the corresponding entropy correction is included, ensuring consistency with the first law of black hole thermodynamics to first order in the EFT expansion.

From Equation~\eqref{eq:hawktempcor}, one finds that for negative $c_6$ the bracket can vanish at finite mass. Formally setting the leading-order correction to zero defines a characteristic mass scale
\begin{align}
    M_{crit}^{-} \equiv \left(2 \pi |c_6|\right)^{\frac{1}{4}} M_p,
\end{align}
at which the first-order approximation to the Hawking temperature vanishes. The temperature reaches a global maximum at
\begin{align}
    M_T^- \equiv \left(10\pi |c_6| \right)^{\frac{1}{4}} M_p.
\end{align}

We stress that the vanishing of $T_H$ at $M_{crit}^{-}$ arises from the truncation of the perturbative expansion and should not be interpreted as a statement about the exact temperature of the underlying quantum-corrected geometry. Rather, it indicates a qualitative freeze-out of the evaporation process within the domain where the EFT corrections become comparable to the leading semiclassical term.

\begin{figure}[ht]
            \centering
            \includegraphics[width=0.8\textwidth]{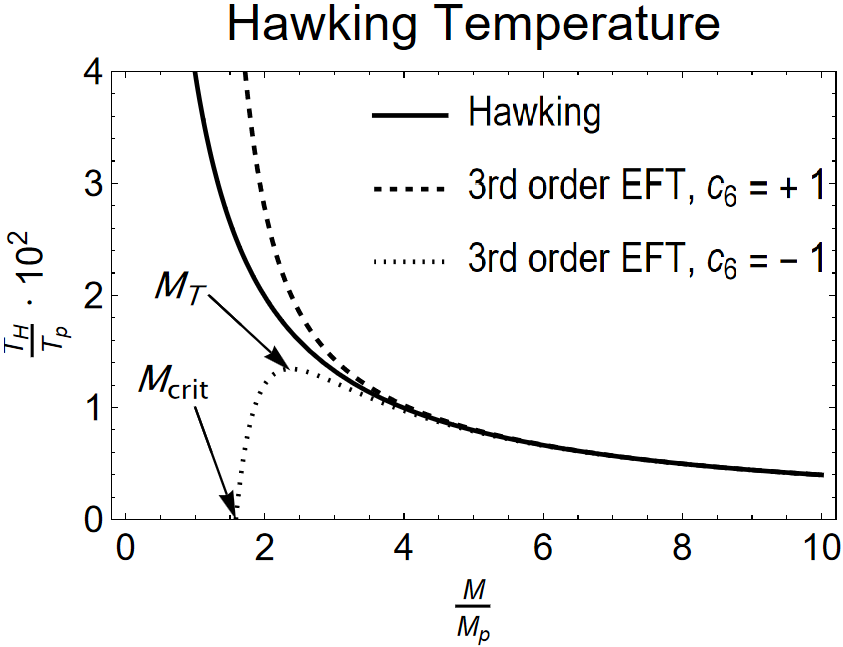}
            \caption{Hawking temperature as a function of the normalized mass $\frac{M}{M_p}$ for three cases: the standard Hawking result ($c_6 = 0$), and third-order effective field theory with $c_6 = +1$ and $c_6 = -1$. For $c_6 < 0$, the temperature reaches a maximum at the scale $M_T^{-}$ and subsequently decreases toward zero as $M \rightarrow M_{crit}^{-}$, indicating evaporation freeze-out.} 
            \label{fig:HawkTemp}
\end{figure}

For positive $c_6$, the bracket in Equation~\eqref{eq:hawktempcor} remains strictly positive, and the temperature never vanishes at this order. A comparison with the standard Hawking result is shown in Figure~\ref{fig:HawkTemp}.

\subsection{Mass-Loss Rate and Freeze-Out Behavior for \texorpdfstring{$c_6 > 0$}{}}

We now consider the case $c_6 > 0$, for which the Hawking temperature remains nonzero but the corrected mass-loss rate exhibits a zero within the truncated expansion.

Starting from the Stefan–Boltzmann law,
\begin{align}\label{eq:stefanboltzmann}
    \frac{d M}{dt} = -\frac{\sigma}{c^2} A_H T_H^4,
\end{align}
with $\sigma$ the Stefan-Boltzmann constant, and inserting the corrected expressions for $A_H$ and $T_H$, we obtain
\begin{align}\label{eq:correctedDiffEq}
    \frac{d M}{dt} = - &\frac{1}{15360\pi} \frac{M_p}{t_p} \left(\frac{M_p}{M}\right)^2 \left[1 - 2\pi c_6 \left(\frac{M_p}{M}\right)^4 \right] + \mathcal{O}\left(c_6^2 \left(\frac{M_p}{M}\right)^{10}\right). 
\end{align}

For $c_6>0$, the leading-order mass-loss rate vanishes at
\begin{align}
    M_{crit}^{+} \equiv (2\pi c_6)^{1/4} M_p,
\end{align}
signaling evaporation freeze-out behavior at this order of the expansion.

As in the $c_6 < 0$ case where the temperature tends to zero, the vanishing here arises from a cancellation between the leading Hawking term and the first EFT correction, and should therefore be interpreted as a feature of the truncated polynomial approximation rather than an exact statement about the full quantum theory.

\begin{figure}[h]
            \centering
            \includegraphics[width=0.8\textwidth]{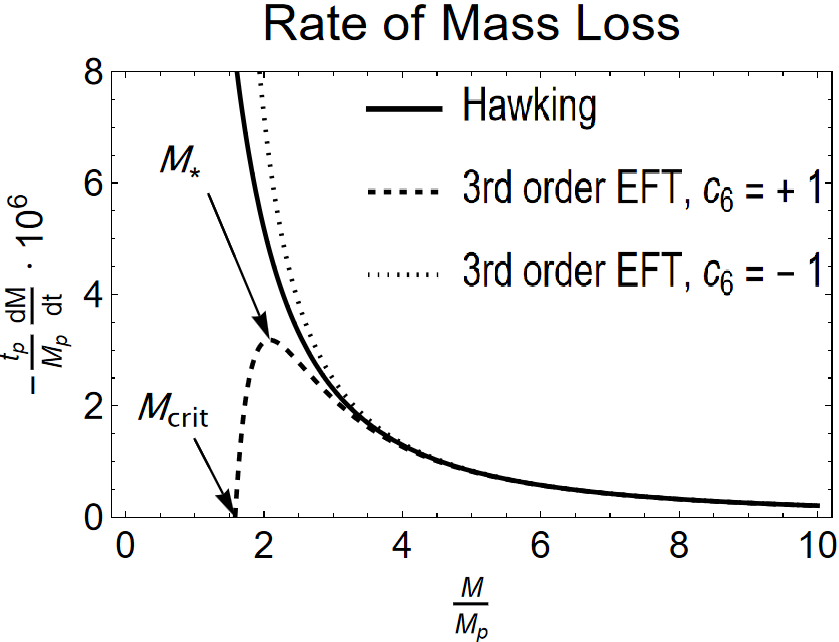}
            \caption{Comparison of the evaporation rate in third-order EFT for positive and negative $c_6$ as a function of the normalized mass $\frac{M}{M_p}$, shown alongside Hawking’s evaporation law. } 
            \label{fig:EvapLawPaperCompareNegPos}
\end{figure}

The evaporation rate also exhibits a local maximum at
\begin{align}
    M_{*} \equiv (6\pi c_6)^{1/4} M_p,
\end{align}
with the ratio $M_{*}/M_{crit}=3^{1/4}$ independent of $c_6$. For $M<M_{*}$ the evaporation rate decreases, in qualitative contrast to the monotonic increase predicted by Hawking. This behavior is illustrated in Figure~\ref{fig:EvapLawPaperCompareNegPos}.

\subsection{Interpretation of the leading-order zero in the evaporation rate \texorpdfstring{\\}{}}

In semiclassical evaporation the emitted power $\frac{dM}{dt}$ scales as $T_H^4$, so that $T_H \to 0$ implies $\frac{dM}{dt}\to 0$. In the present case, however, the mass-loss rate in Equation~\eqref{eq:correctedDiffEq} is obtained by expanding the corrected temperature and horizon area to first order in the dimensionless parameter $\epsilon \sim c_6 (M_p/M)^4$. The resulting expression is therefore a truncated polynomial in $(M_p/M)^4$.

The critical mass $M_{crit}^{+}$ arises from a cancellation between the leading Hawking contribution and the first EFT correction within this truncated expansion. The vanishing of $\frac{dM}{dt}$ at this order should thus be understood as signaling that the correction has become comparable to the leading term, rather than as a proof that radiation ceases in the exact theory. Once higher-order curvature operators are included, additional contributions of order $(M_p/M)^{10}$ and beyond will modify the structure of the polynomial and shift the location of this root, or potentially remove it altogether.

The important physical point is therefore not the literal existence of a zero in the truncated mass-loss rate, but the qualitative change in behavior: below a characteristic mass scale, the evaporation rate no longer increases monotonically as in Hawking’s result, but instead exhibits a maximum and subsequently decreases. This signals a dynamical slow-down of the evaporation process within the perturbative framework.

\section{Thermodynamic Properties of the Freeze-Out Regime}\label{sec:thermo}
Having identified the emergence of a characteristic mass scale at which the evaporation dynamics change qualitatively, we now examine the associated thermodynamic structure. We begin by analyzing the time-dependent evolution of the Hawking temperature, which provides a direct probe of how the system heats up or cools during evaporation. We then turn to the evaporation time and the heat capacity, which respectively characterize the dynamical and thermodynamic aspects of the freeze-out behavior.

\subsection{Temperature Evolution}\label{sec:tempev}
A useful diagnostic of the evaporation process is the evolution of the Hawking temperature with time. Using the chain rule, we can write
\begin{equation}
    \frac{dT_H}{dt} = \frac{dT_H}{dM}\frac{dM}{dt}.
\end{equation}
This relation follows directly from treating the temperature as a function of the mass, $T_H = T_H(M(t))$, and remains valid within the perturbative effective field theory framework.

Substituting the corrected expressions for the Hawking temperature and the mass-loss rate, Equations~\eqref{eq:hawktempcor} and \eqref{eq:correctedDiffEq}, and expanding consistently to first order in $c_6 (M_p/M)^4$, one obtains
\begin{equation}
    \frac{dT_H}{dt} = \frac{1}{122880\pi^2}\left(\frac{T_p}{t_p}\right)\left(\frac{M_p}{M}\right)^4 \left[1 + 8\pi c_6 \left(\frac{M_p}{M}\right)^4 \right] + \mathcal{O}\left(c_6^2 \left(\frac{M_p}{M}\right)^{12}\right).
\end{equation}

This expression reveals a qualitative difference between positive and negative values of $c_6$. For $c_6>0$, the bracket remains strictly positive, and therefore
\begin{equation}
    \frac{dT_H}{dt} > 0
\end{equation}
for all masses within the regime of validity of the expansion. This is consistent with the standard picture of Hawking evaporation: as time progresses and the black hole loses mass, its temperature increases.

For $c_6<0$, however, the correction term enters with the opposite sign. As the mass decreases and $(M_p/M)^4$ grows, the bracket can vanish at a finite mass scale, defined by
\begin{equation}
    1 + 8\pi c_6 \left(\frac{M_p}{M}\right)^4 = 0.
\end{equation}
At this point, $\frac{dT_H}{dt} = 0$, and for smaller masses one finds $\frac{dT_H}{dt} < 0$. The temperature therefore reaches a maximum and subsequently decreases as the evaporation proceeds.

The existence of a zero in $\frac{dT_H}{dt}$ for $c_6<0$ implies the presence of a maximal Hawking temperature. This is in agreement with the behavior already visible in the explicit expression for $T_H(M)$ in Equation~\eqref{eq:hawktempcor}, which exhibits a maximum at the scale $M_T^{-}$.

It is important to note, however, that the location of this maximum does not coincide exactly with the one inferred from $\frac{dT_H}{dt} = 0$. The former yields $M_T^{-} = (10\pi |c_6|)^{1/4} M_p$, while the latter leads to a slightly different numerical coefficient. This discrepancy is a consequence of the truncation of the perturbative expansion: both $T_H$ and $\frac{dT_H}{dt}$ are computed consistently only up to first order in $c_6 (M_p/M)^4$, and higher-order terms would shift the precise location of the maximum.

The physically robust conclusion is therefore the qualitative one: for $c_6<0$, the temperature evolution becomes non-monotonic, with a maximum followed by a cooling phase, whereas for $c_6>0$ the temperature increases monotonically throughout the evaporation process.

\subsection{Evaporation Time}
Another useful diagnostic of the late-stage dynamics is the evaporation time required for a black hole to evolve toward the characteristic mass scale $M_{\rm crit}$. In standard Hawking evaporation, the mass-loss rate scales as $\frac{dM}{dt} \propto -M^{-2}$, and the evaporation time is defined as the time required for a black hole of initial mass $M_0$ to reach zero mass,
\begin{equation}
    t_{\rm evap}(M_0) = \int_{0}^{M_0} \frac{dM}{|dM/dt|} \propto M_0^3,
\end{equation}
which is finite.

In the present effective field theory framework, however, the evaporation dynamics differ qualitatively.

\vspace{4mm}

For $c_6>0$, the corrected mass-loss rate does not remain monotonic but instead develops a local maximum and subsequently decreases as the mass approaches the characteristic scale $M_{\rm crit}$. In this case, the leading-order mass-loss rate vanishes at $M_{\rm crit}$, and the physically relevant quantity is no longer the time required to reach zero mass, but rather the time required for a black hole of mass $M$ to approach the freeze-out scale $M_{\rm crit}$ up to some small $\delta > 0$ where $M_{\rm crit} + \delta < M$,
\begin{equation}\label{eq:inttimeEFT}
    \tau(M) = \int_{M_{\rm crit} + \delta}^{M} \frac{dM'}{|dM'/dt|}.
\end{equation}

The behavior of this integral can be understood analytically. Within the truncated effective field theory expression, the mass-loss rate defined through Equation \eqref{eq:correctedDiffEq} is smooth and exhibits a simple zero at $M = M_{\rm crit}$, defined by the vanishing of the leading-order correction term. In particular, the mass-derivative of the mass-loss rate \eqref{eq:correctedDiffEq} at $M_{crit}$ is nonzero, so that the zero is of first order. Near the critical mass, the mass-loss rate therefore admits a linear expansion,
\begin{equation}\label{eq:linear}
    \frac{dM}{dt} \approx - \kappa \, (M - M_{\rm crit}),
\end{equation}
where $\kappa = \left.\frac{d}{dM}\left(-\frac{dM}{dt}\right)\right|_{M_{\rm crit}} > 0$ is determined by the first derivative of the evaporation law. Substituting this form into Equation \eqref{eq:inttimeEFT} yields
\begin{equation}
    \tau(M) = \int_{M_{\rm crit} + \delta}^{M} \frac{dM'}{\kappa (M' - M_{\rm crit})} = \frac{1}{\kappa} \ln{|M-M_{\rm crit}|} - \frac{1}{\kappa} \ln{|\delta|}.
\end{equation}

Taking the limit $\delta \downarrow 0$ then gives
\begin{equation}
    \lim_{\delta \downarrow 0} \tau(M) = \infty,
\end{equation}
so that the evaporation time diverges logarithmically as $M \to M_{\rm crit}$.

The logarithmic divergence reflects a slow-down associated with the appearance of a simple zero in the mass-loss rate. As long as the leading-order evaporation law develops a first-order root, the approach to the critical mass is necessarily asymptotic.

The linearized Equation \eqref{eq:linear} is readily solved:
\begin{equation}
    M(t) - M_{\rm crit} = (M_0 - M_{\rm crit}) e^{-\kappa t},
\end{equation}
where $M_0$ is the mass at some reference time $t = 0$. Therefore, in the vicinity of the critical mass, $M(t)$ approaches the critical value exponentially slowly in time, and the critical mass is never reached in finite time within the truncated dynamics.

\vspace{4mm}

For negative $c_6$, the structure of the late-stage dynamics differs qualitatively from the $c_6>0$ case. As discussed in Section~\ref{sec:evaporation}, the corrected Hawking temperature reaches a maximum at $M_T^{-} = (10\pi |c_6|)^{1/4} M_p$ and subsequently decreases as the mass decreases further, vanishing at the characteristic scale $M_{\rm crit}$. In contrast to the $c_6>0$ case, however, the mass-loss rate does not develop a zero at $M_{\rm crit}$ within the truncated expansion.

The evaporation time required for a black hole of initial mass $M_0$ to reach the critical mass can therefore be computed directly from the corrected evaporation law. Using Equation~\eqref{eq:correctedDiffEq}, one finds to first order in $c_6 (M_p/M)^4$:
\begin{align}
    \tau(M_0 \rightarrow M_{\rm crit}) &= \int_{M_{\rm crit}}^{M_0} \frac{dM'}{|dM'/dt|} \nonumber \\
    &= \int_{M_{\rm crit}}^{M_0} 15360\pi \left(\frac{t_p}{M_p}\right) \left(\frac{M'}{M_p}\right)^2 \left(1 - 2 \pi |c_6|\left(\frac{M_p}{M'}\right)^4 \right) dM' \\
    %&= 5120\pi t_p \left(\frac{M_0^3 - M_{\rm crit}^3}{M_p^3} + 6\pi |c_6| \left(\frac{M_p}{M_0} - \frac{M_p}{M_{\rm crit}}\right)\right).\\
    &= 5120\pi t_p \frac{M_0^3 - M_{\rm crit}^3}{M_p^3} \left(1 - 6\pi |c_6| \frac{M_p^4}{M_o M_{\rm crit}\left(M_0^2 + M_0 M_{\rm crit} + M_{\rm crit}^2\right)}\right).
\end{align}

\begin{figure}[ht]
            \centering
            \includegraphics[width=0.8\textwidth]{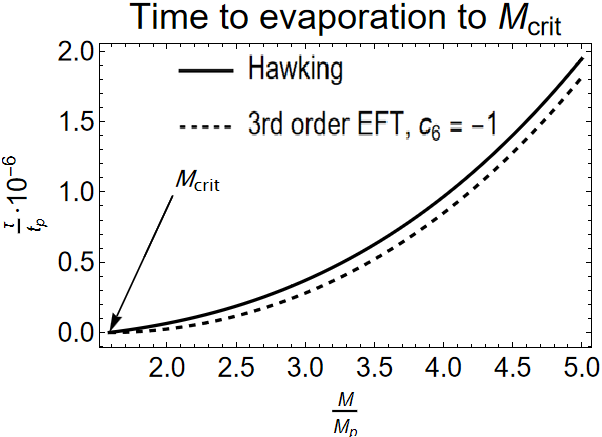}
            \caption{Comparison of the time to evaporation from an initial mass $M$ to the critical mass $M_{crit}$ as a function of the normalized mass $\frac{M}{M_p}$ of a Schwarzschild black hole in third-order effective field theory for negative $c_6$, together with Hawking’s prediction.}
            \label{fig:EvapTime}
\end{figure}

The first term reproduces the classical Schwarzschild evaporation time, while the second term encodes the leading higher-curvature correction. Since the integrand remains finite throughout the integration range, the evaporation time to reach $M_{\rm crit}$ is finite. A comparison with the standard Hawking result is shown in Figure~\ref{fig:EvapTime}.

The behavior at the critical mass itself requires some care. Within the truncated expansion, the Hawking temperature vanishes at $M_{\rm crit}$, while the mass-loss rate remains finite and negative. At the same time, the temperature evolution discussed in Section~\ref{sec:tempev} shows that $\frac{dT_H}{dt}$ remains negative at this point. Taken at face value, this would imply that the temperature would continue to decrease beyond $T_H=0$, which is unphysical.

This apparent inconsistency reflects the breakdown of the perturbative expansion near $M \sim M_{\rm crit}$. The condition $T_H=0$ arises from a cancellation between the leading Hawking contribution and the first higher-curvature correction in Equation~\eqref{eq:hawktempcor}. In this regime, the expansion parameter $\epsilon \sim c_6 (M_p/M)^4$ becomes of order unity, and higher-order terms that have been neglected are expected to modify both $T_H$ and $\frac{dM}{dt}$.

The physically robust conclusion is therefore limited to the approach toward the critical scale: for $c_6<0$, the black hole evolves to $M \sim M_{\rm crit}$ in a finite time, while the temperature decreases toward zero after reaching a maximum at $M_T^{-}$. The truncated dynamics do not reliably determine the subsequent evolution. In particular, whether the system settles into a long-lived remnant, continues to evaporate with a modified temperature, or undergoes a qualitatively different transition depends on higher-order corrections beyond the present EFT truncation.

Nevertheless, the analysis demonstrates that, in contrast to the $c_6>0$ case where evaporation slows down asymptotically, the $c_6<0$ scenario does not exhibit a dynamical freeze-out at the level of the mass-loss rate. Instead, it is characterized by a finite-time approach to a regime in which the Hawking temperature vanishes and the perturbative description itself ceases to be reliable.

\subsection{Heat Capacity and Local Thermodynamic Structure}
A complementary diagnostic of the late-stage behavior of a black hole is provided by the heat capacity,
\begin{equation}
    C \equiv \frac{dM}{dT_H}.
\end{equation}
For a classical Schwarzschild black hole one has $C < 0$ for all $M$, enforcing runaway evaporation \cite{Hawking1975}.

Since the corrected Hawking temperature and Wald entropy satisfy the first law $dM=T_HdS$ to first order in the EFT expansion \cite{Calmet2021Entropy}, the heat capacity may equivalently be written as
\begin{align}
    C = T_H \frac{dS}{dT_H},
\end{align}
which is the form commonly used in black-hole thermodynamics. The corresponding entropy $S$ is the corrected Wald entropy derived in \cite{Calmet2021Entropy}, ensuring consistency between the geometric and thermodynamic descriptions at this order in the EFT expansion.

In the third-order EFT considered here, the heat capacity becomes
\begin{align}\label{eq:heatcap}
    C = -8\pi \left(\frac{M_p}{T_p} \right) \left(\frac{M}{M_p}\right)^2 \left[1 + 10\pi c_6 \left(\frac{M_p}{M}\right)^4 \right] + \mathcal{O}\!\left(c_6^2 \left(\frac{M_p}{M}\right)^6\right).
\end{align}

The qualitative behavior depends on the sign of $c_6$.

\subsubsection{Case \texorpdfstring{$c_6 > 0$\\}{}}
\noindent The heat capacity remains negative for all masses within the perturbative regime. However, its magnitude no longer grows monotonically as $M$ decreases. Instead, it exhibits a local maximum at
\begin{equation}
M_T^{+} = (10\pi c_6)^{1/4} M_p.
\end{equation}
Below this scale, $C$ decreases and the system therefore remains thermodynamically unstable in the canonical sense ($C<0$).

\subsubsection{Case \texorpdfstring{$c_6 < 0$\\}{}}

\noindent For negative $c_6$, the heat capacity changes sign at
\begin{equation}
    M_T^{-} = (10\pi |c_6|)^{1/4} M_p.
\end{equation}
Below this mass scale one finds $C>0$ within the truncated expansion.

Since Equation \eqref{eq:heatcap} is analytic throughout the perturbative regime, the heat capacity remains continuous at $M_T^{-}$. At this mass the derivative $dT_H/dM$ vanishes, but because we consider $C=dM/dT_H$ consistently to first order in $c_6(M_p/M)^4$, $C$ passes smoothly through zero and changes sign. No divergence occurs at this order, and the transition is therefore continuous rather than singular.

A positive heat capacity indicates that the temperature decreases as the mass decreases, in qualitative contrast with the classical Schwarzschild behavior. This sign change is directly connected to the presence of a maximum in the corrected Hawking temperature and allows for $T_H$ to decrease with decreasing $M$, behavior which is impossible for $C<0$.

The appearance of $C>0$ does not imply stability in a strict thermodynamic sense, because the analysis neglects fluctuations and backreaction beyond leading order. However, it does indicate a departure from Hawking's heat capacity predictions.

The vanishing of the Hawking temperature at $M_{\rm crit}^{-}$ does not imply that the heat capacity vanishes. Instead, Equation \eqref{eq:heatcap} shows that $C$ remains finite and positive. This reflects the fact that $dS/dT_H$ diverges sufficiently rapidly to compensate for the vanishing temperature, consistent with the first law.

The sign change of the heat capacity separates the familiar Schwarzschild-like regime with negative heat capacity from a low-mass regime in which the truncated EFT predicts $C>0$. While this resembles the change of thermodynamic behavior discussed in various quantum-gravity inspired black-hole models, the absence of any divergence in the heat capacity indicates that the present EFT does not exhibit a genuine second-order thermodynamic phase transition within the perturbative approximation. Rather, the sign change should be interpreted as a smooth crossover between two local thermodynamic regimes.

\begin{figure}[ht]
            \centering
            \includegraphics[width=0.8\textwidth]{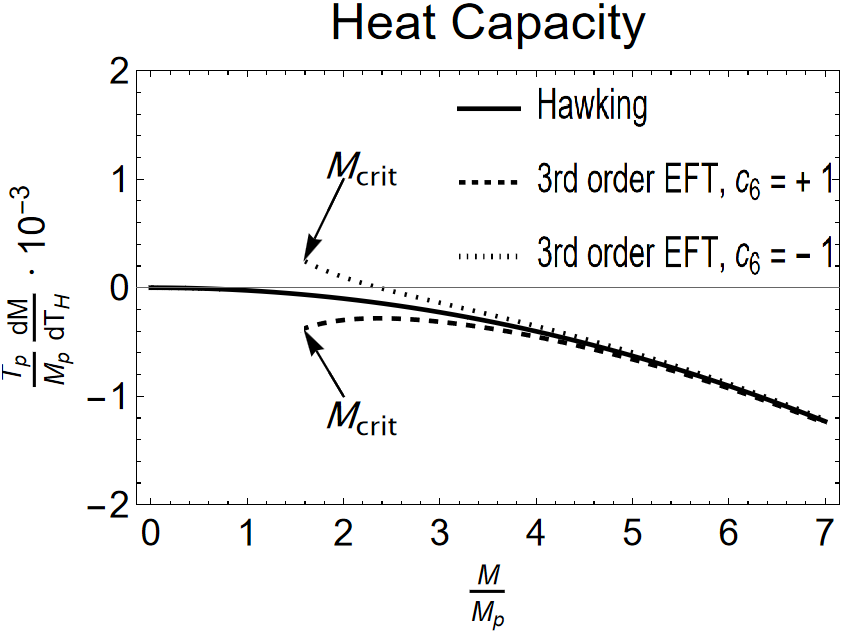}
            \caption{Comparison of the heat capacity $C \equiv \frac{dM}{dT_H}$ as a function of the normalized mass $\frac{M}{M_p}$ of a Schwarzschild black hole in third-order effective field theory for positive and negative $c_6$, together with Hawking’s prediction. }
            \label{fig:HeatCap1}
\end{figure}

Figure~\ref{fig:HeatCap1} compares the heat capacity for both signs of $c_6$ with the classical Schwarzschild result. The behavior of the heat capacity mirrors that of the corrected Hawking temperature shown in Figure~\ref{fig:HawkTemp}, providing a consistent thermodynamic interpretation of the evaporation slow-down and freeze-out behavior within the truncated effective field theory.

\subsection{Infrared Consistency and Planckian Regime}

For $M \gg M_p$, all corrections proportional to $\epsilon \sim c_6 (M_p/M)^4$ are strongly suppressed. The standard Schwarzschild scalings,
\begin{equation}
    T_H \propto M^{-1},
    \qquad
    C \propto -M^2,
    \qquad
    \frac{dM}{dt} \propto -M^{-2},
\end{equation}
are recovered, ensuring consistency with the infrared limit of general relativity.

Deviations become significant only when $c_6 (M_p/M)^4 \sim \mathcal{O}(1)$, i.e. precisely in the regime where the perturbative hierarchy begins to break down. The qualitative modifications identified above (emergence of a characteristic mass scale, non-monotonic evaporation rate, and sign-dependent behavior of the heat capacity) therefore represent consequences of the leading higher-curvature correction, even though the exact endpoint of evaporation cannot be determined within the truncated expansion.

Taken together, the analysis demonstrates that cubic curvature operators can qualitatively alter late-stage black hole thermodynamics, inducing evaporation slow-down and modifying the behavior of the heat capacity. The precise fate of the system, however, requires control over the full derivative expansion, to which we now turn.

\section{Regime of Validity of the Effective Field Theory}\label{sec:validity}
The analysis of Sections~\ref{sec:evaporation} and \ref{sec:thermo} demonstrates that cubic curvature corrections can qualitatively modify late-stage black hole evaporation within a truncated effective field theory. In particular, the appearance of a characteristic mass scale $M_{\mathrm{crit}}$ signals a dynamical slow-down of the evaporation process and a corresponding change in the thermodynamic behavior of the system. 

Before assigning physical significance to these features, it is necessary to determine whether the regime in which they arise lies within the domain of validity of the derivative expansion. Since gravitational effective field theory is organized as a local expansion in curvature invariants suppressed by the Planck scale \cite{Donoghue1994Leading,Burgess2004Everyday,Donoghue2012Effective}, its predictive control relies on a parametric hierarchy between successive higher-curvature operators. In this section we examine this hierarchy explicitly. 

We first evaluate the Kretschmann curvature invariant at the horizon to determine whether the freeze-out scale corresponds to sub-Planckian or Planckian curvature. We then compare the cubic contribution to generic quartic and higher-order operators in the action to assess whether the truncation remains self-consistent near $M_{\mathrm{crit}}$.

\subsection{Curvature Invariants and EFT Validity at the Critical Mass}\label{sec:kretschmann}
The appearance of a characteristic freeze-out mass $M_{\mathrm{crit}}$ in the corrected evaporation dynamics raises a central question: does this scale lie within the regime of validity of the truncated effective theory? To answer this, we directly evaluate the curvature at the horizon and compare it to the cutoff scale at $M \sim M_{\mathrm{crit}}$.

In units $c=\hbar=1$, the cutoff is set by the reduced Planck mass $M_p$ or equivalently the Planck length $\ell_p$. Among the various invariants that can be constructed, the Kretschmann scalar provides a particularly useful diagnostic because it measures the full magnitude of spacetime curvature and remains nonzero even in vacuum solutions where the Ricci scalar vanishes such as the classical Schwarzschild solution. It therefore captures the relevant local curvature scale governing the validity of the derivative expansion. The derivative expansion is reliable only when curvature invariants satisfy
\begin{equation}
    \mathcal{K} \equiv R_{\mu\nu\rho\sigma}R^{\mu\nu\rho\sigma} \ll \ell_p^{-4}.
\end{equation}

Because Hawking radiation is determined by near-horizon physics, the appropriate quantity to examine is the curvature evaluated at the Killing horizon $r=r_H$. For a classical Schwarzschild black hole one finds
\begin{equation}
    \mathcal{K}_{\rm SS}(r_H) = \frac{3}{4}\frac{1}{r_H^4}
    \sim \left(\frac{M_p}{M}\right)^4 \ell_p^{-4},
\end{equation}
while the Ricci scalar vanishes identically in vacuum \cite{Wald1984,Carroll2004}. Thus, classical curvature remains parametrically small only when $M \gg M_p$.

Including the cubic correction, the metric receives perturbative modifications controlled by the dimensionless expansion parameter
\begin{equation}
    \epsilon \sim c_6 \left(\frac{M_p}{M}\right)^4.
\end{equation}
To first order in $\epsilon$, curvature invariants at the horizon acquire corrections of relative order $\epsilon$, and the perturbative description remains self-consistent only for $|\epsilon| \ll 1$.

It is worth emphasizing that the perturbative expansion employed throughout this work is organized in powers of the dimensionless parameter
\begin{equation}
    \epsilon \sim c_6\left(\frac{M_p}{M}\right)^4,
\end{equation}
rather than in powers of the Wilson coefficient $c_6$ itself. Consequently, Wilson coefficients of natural size, $c_6=\mathcal{O}(1)$, are entirely compatible with the effective field theory description provided that the curvature remains sufficiently small so that $|\epsilon|\ll1$. Conversely, even if $|c_6|\ll1$, the perturbative expansion necessarily breaks down once $|\epsilon|\sim1$, where higher-curvature operators become equally important. The regime of validity of the analysis is therefore controlled by the combination $c_6(M_p/M)^4$, not by the magnitude of $c_6$ in isolation.

The freeze-out mass derived in Section~\ref{sec:evaporation} is defined precisely by the condition
\begin{equation}
    |c_6| \left(\frac{M_p}{M_{\mathrm{crit}}}\right)^4 \sim \mathcal{O}(1),
\end{equation}
which corresponds to $|\epsilon| \sim 1$. Evaluating the curvature at this scale yields
\begin{equation}
    \mathcal{K}(r_H) \sim \left(\frac{M_p}{M}\right)^4 \ell_p^{-4}
    \sim \frac{1}{|c_6|}\,\ell_p^{-4}
    \qquad \text{at } M = M_{\mathrm{crit}}.
\end{equation}
For Wilson coefficients $|c_6| = \mathcal{O}(1)$, the Kretschmann scalar is therefore Planckian at the horizon, indicating that the derivative expansion has reached the boundary of its validity.

This conclusion is insensitive to the precise value of the Wilson coefficient. If $|c_6| \ll 1$, the critical mass lies below the Planck scale and the curvature becomes even larger,
\begin{equation}
    \mathcal{K}(M_{\mathrm{crit}}) \sim \frac{1}{|c_6|}\,\ell_p^{-4} \gg \ell_p^{-4},
\end{equation}
placing the solution far outside the EFT regime. Conversely, taking $|c_6| \gg 1$ pushes $M_{\mathrm{crit}}$ far above $M_p$, but does not extend the range of validity of the effective theory. In gravitational EFT, parametrically large Wilson coefficients signal the onset of strong coupling of higher-curvature interactions and effectively lower the cutoff scale of the derivative expansion rather than enlarging it \cite{Burgess2004Everyday,Donoghue2012Effective}.

We therefore conclude that there exists no parametric regime in which the freeze-out mass lies safely within the perturbative domain of the local curvature expansion. The apparent slow-down of evaporation arises precisely as the dimensionless expansion parameter approaches unity and curvature at the horizon becomes Planckian. The result should thus be interpreted as the EFT being driven to the boundary of its applicability, rather than as a controlled prediction deep within the perturbative regime.

\vspace{3mm}

\subsection{Parametric Comparison with Quartic and Higher--Order Curvature Operators}
The characteristic mass
\begin{equation}
    M_{\mathrm{crit}} \sim (2\pi |c_6|)^{1/4} M_p
\end{equation}
coincides parametrically with the onset of Planckian curvature at the horizon, as established in the previous subsection. This observation motivates a complementary question: even if curvature itself is already approaching the cutoff, does the cubic operator remain parametrically dominant within the effective action near this scale, or does the entire derivative expansion lose its hierarchical organization?

The local gravitational effective action is organized as an expansion in curvature invariants with increasing mass dimension \cite{Donoghue1994Leading,Burgess2004Everyday}. Retaining only the leading nontrivial vacuum correction, the action used in this work is
\begin{equation}
    S_{EFT,L} = \int d^4 x \sqrt{-g} \left(\frac{M_p^2}{16\pi}R + c_1 R^2 + c_2 R_{\mu \nu} R^{\mu \nu} + c_3 R_{\mu \nu \rho \sigma} R^{\mu \nu \rho \sigma} + \frac{c_6}{M_p^2} R^{\mu \nu}{}_{\alpha \sigma} R^{\alpha \sigma}{}_{\delta \gamma}R^{\delta \gamma}{}_{\mu \nu} + ...\right), 
\end{equation}
where the $c_i$ are dimensionless Wilson coefficients and the ellipsis denotes higher–order operators. Quadratic curvature terms do not modify the Schwarzschild solution in vacuum, so the cubic invariant constitutes the first nontrivial correction \cite{Calmet2021Entropy, LuPerkinsPopeStelle2015, Calmet2018Quantum, Calmet2018VanishingCurvature}.

At the next order in the derivative expansion, one generically expects quartic curvature operators of schematic form
\begin{equation}
    S^{(4)} \sim \int d^4x \sqrt{-g}\; \frac{c_8}{M_p^4} \, \mathcal{R}^4,
\end{equation}
where $\mathcal{R}^4$ denotes contractions of four Riemann tensors and $c_8$ is dimensionless. Such operators arise in perturbative quantum gravity at higher loop order \cite{Goroff1986Ultraviolet,VandeVen1992}, and more generally follow from EFT power counting \cite{Donoghue1994Leading,Donoghue2012Effective}.

Rather than solving the full higher–order corrected field equations, we perform a parametric comparison based on curvature scaling at the horizon. Working in natural units $c=\hbar=1$, the only dimensionful scale near the Schwarzschild horizon is the horizon radius
\begin{equation}
    r_H \sim \frac{2M}{M_p^2},
\end{equation}
so that curvature tensors scale parametrically as
\begin{equation}
    R_{\mu\nu\rho\sigma} \sim \frac{1}{r_H^2} \sim \frac{M_p^4}{M^2}.
\end{equation}

Since each operator contributes to the equations of motion with two derivatives acting on the metric per Riemann tensor, the relative scaling of operators in the Lagrangian density matches their relative scaling in the field equations. Evaluated parametrically at the horizon, the Einstein–Hilbert term scales as
\begin{equation}
    \mathcal{L}_{\rm EH} \sim M_p^2 R \sim M_p^2 \frac{M_p^4}{M^2} = \frac{M_p^6}{M^2}.
\end{equation}

The cubic operator contributes
\begin{equation}
    \mathcal{L}_{\rm cubic} \sim \frac{c_6}{M_p^2} \left(\frac{M_p^4}{M^2}\right)^3 = c_6 \frac{M_p^{10}}{M^6}.
\end{equation}
The ratio to the Einstein–Hilbert term is therefore
\begin{equation}
    \frac{\mathcal{L}_{\rm cubic}}{\mathcal{L}_{\rm EH}} \sim c_6 \left(\frac{M_p}{M}\right)^4,
\end{equation}
which reproduces the expansion parameter $\epsilon$ controlling the perturbative solution \cite{Calmet2021Entropy}.

At quartic order,
\begin{equation}
    \mathcal{L}_{\rm quartic} \sim \frac{c_8}{M_p^4} \left(\frac{M_p^4}{M^2}\right)^4 = c_8 \frac{M_p^{12}}{M^8},
\end{equation}
so that
\begin{equation}
    \frac{\mathcal{L}_{\rm quartic}}{\mathcal{L}_{\rm EH}} \sim c_8 \left(\frac{M_p}{M}\right)^6.
\end{equation}

More generally, an operator containing $n \geq 3$ powers of the Riemann tensor has schematic form
\begin{equation}
    S^{(n)} \sim \int d^4x \sqrt{-g}\; \frac{c_{2n}}{M_p^{2n-4}}\,\, \mathcal{R}^{n},
\end{equation}
where $c_{2n}$ is dimensionless. Parametrically,
\begin{equation}
    \frac{\mathcal{L}_{n}}{\mathcal{L}_{\rm EH}}
    \sim
    c_{2n}
    \left(\frac{M_p}{M}\right)^{2n-2}.
\end{equation}

We now evaluate these expressions at the characteristic mass defined by
\begin{equation}
    |c_6|\left(\frac{M_p}{M_{\rm crit}}\right)^4 \sim 1.
\end{equation}
At this scale,
\begin{equation}
    \left(\frac{M_p}{M_{\rm crit}}\right)^{2n-2}
    \sim
    |c_6|^{-\frac{n-1}{2}},
\end{equation}
so that
\begin{equation}
    \frac{\mathcal{L}_n}{\mathcal{L}_{\rm EH}}
    \sim
    c_{2n}\,
    |c_6|^{-\frac{n-1}{2}}.
\end{equation}

Unless higher–order Wilson coefficients are parametrically suppressed relative to $c_6$, operators of all orders in the derivative expansion become comparable to the Einstein–Hilbert term when $M \sim M_{\rm crit}$. In particular, quartic curvature invariants are generically of order unity at the same scale where the cubic operator ceases to be perturbative.

This scaling argument shows that the scale $M_{\rm crit}$ identified in Section~\ref{sec:evaporation} marks not merely the breakdown of the cubic truncation, but the loss of parametric hierarchy of the entire local derivative expansion. The apparent freeze–out behavior therefore coincides with the regime in which curvature becomes Planckian and higher–curvature operators of all orders contribute at comparable strength.

Consequently, the slow–down of evaporation at $M \sim M_{\rm crit}$ should be interpreted as the EFT approaching its cutoff, rather than as a controlled prediction of cubic gravity alone. Within the regime $|c_6|(M_p/M)^4 \ll 1$, the hierarchy of the derivative expansion remains intact; once this condition fails, no finite truncation of the local curvature expansion can be regarded as parametrically reliable.

\section{Extensions}\label{sec:extensions}
The analysis presented in this work has focused on Schwarzschild black holes and on a simplified evaporation model based on the Stefan–Boltzmann approximation. In this section we briefly discuss two natural extensions: (i) the inclusion of greybody factors in the evaporation law, and (ii) the qualitative implications of cubic curvature corrections for charged and rotating black holes. The purpose of this section is not to provide complete derivations, but rather to assess whether the freeze-out behavior identified above is structurally robust.

\subsection{Greybody Corrections}\label{sec:greybody}
Throughout this work, the evaporation rate has been modeled using a Stefan–Boltzmann–type approximation of the schematic form
\begin{equation}
    \frac{dM}{dt} \sim A_H\, T_H^4,
\end{equation}
where $A_H$ is the horizon area and $T_H$ the corrected Hawking temperature. While this captures the leading scaling with temperature and horizon area, it assumes that the black hole radiates as an ideal blackbody. In reality, the emitted radiation is partially filtered by the curved spacetime outside the horizon, leading to frequency-dependent deviations from a perfect thermal spectrum.

These deviations are encoded in so-called greybody factors, which arise from the scattering of field modes in the effective potential surrounding the black hole. They quantify the probability that radiation produced near the horizon escapes to infinity rather than being reflected back.

The full semiclassical emission rate is given by \cite{Page1976a,Page1976b}
\begin{equation}
\label{eq:fullflux}
    \frac{dM}{dt} = - \sum_{s,\ell} \left(2 \ell + 1\right)\int_0^\infty \frac{\Gamma_{s\ell}(\omega)} {e^{\omega/T_H} \pm 1} \, \frac{\omega\, d\omega}{2\pi},
\end{equation}
where $s$ labels the spin of the emitted field, $\ell$ the angular momentum mode, and $\Gamma_{s\ell}(\omega)$ are the greybody factors. The factors $\Gamma_{s\ell}(\omega)$ encode the probability that a mode produced near the horizon propagates through the curvature–induced potential barrier to reach asymptotic infinity. Reviews of the formalism may be found in \cite{FabbriNavarro2005}.

To assess whether the freeze--out behavior obtained in the Stefan--Boltzmann--blackbody approximation persists in the presence of greybody effects, it is useful to rewrite Equation~\eqref{eq:fullflux} in a form that isolates the dependence on the horizon temperature and radius. Introducing the dimensionless variable
\begin{equation}
    x \equiv \frac{\omega}{T_H},
\end{equation}
so that $\omega = x T_H$ and $d\omega = T_H\,dx$, Equation~\eqref{eq:fullflux} becomes
\begin{equation}
    \frac{dM}{dt} = -\frac{T_H^2}{2\pi} \sum_{s,\ell} \left(2 \ell + 1\right) \int_0^\infty \frac{x\,\Gamma_{s\ell}(xT_H)}{e^{x}\pm1}\,dx .
\end{equation}

For a static spherically symmetric black hole, the greybody factors depend on frequency only through the dimensionless combination $\omega r_H$ \cite{Page1976a,Page1976b,FabbriNavarro2005}. It is therefore convenient to write
\begin{equation}
    \Gamma_{s\ell}(\omega)=\Gamma_{s\ell}(\omega r_H),
\end{equation}
which yields
\begin{equation}
    \frac{dM}{dt} = -\frac{T_H^2}{2\pi} \sum_{s,\ell}\left(2\ell + 1\right) \int_0^\infty \frac{x\,\Gamma_{s\ell}(x T_H r_H)}{e^{x}\pm1}\,dx .
\end{equation}

\noindent It is now useful to define the dimensionless quantity
\begin{equation}
    y \equiv T_H r_H ,
\end{equation}
and the associated dimensionless emission integral
\begin{equation}
\label{eq:Idef}
    \mathcal{I}(y) = \sum_{s,\ell}\left(2 \ell + 1\right) \int_0^\infty \frac{x\,\Gamma_{s\ell}(x y)}{e^{x}\pm1}\,dx .
\end{equation}
In terms of this function the evaporation rate takes the compact form
\begin{equation}
\label{eq:masterflux}
    \frac{dM}{dt} = -\frac{T_H^2}{2\pi}\,\mathcal{I}(y).
\end{equation}

We now evaluate the argument $y=T_H r_H$ using the corrected expressions defined earlier in this work. Using Equations \ref{eq:horradcor} and \ref{eq:hawktempcor}, we find
\begin{equation}
    T_H r_H = \frac{T_p \ell_p}{4\pi} \left[1-3\pi c_6 \left(\frac{M_p}{M}\right)^4\right] + \mathcal{O}\left(c_6^2\left(\frac{M_p}{M}\right)^8\right).
\end{equation}

\noindent In natural units, the leading value becomes
\begin{equation}
    y_0 = \frac{1}{4\pi},
\end{equation}
and we may now expand the function $\mathcal{I}(y)$ around $y_0$. Defining
\begin{align}
    \mathcal{I}_0 \equiv \mathcal{I}(y_0), \qquad
    \mathcal{I}_0' \equiv \left.\frac{d\mathcal{I}}{dy}\right|_{y_0},
\end{align}
a Taylor expansion gives
\begin{align}
    \mathcal{I}(y) &= \mathcal{I}_0 + \mathcal{I}_0'(y-y_0) + \mathcal{O}\left(c_6^2\left(\frac{M_p}{M}\right)^8\right) \\
    &= \mathcal{I}_0 - \frac{3}{4} c_6 \left(\frac{M_p}{M} \right)^4\,\mathcal{I}_0' + \mathcal{O}\left(c_6^2\left(\frac{M_p}{M}\right)^8\right).
\end{align}

Importantly, both $\mathcal{I}_0$ and $\mathcal{I}_0'$ are pure numbers. This follows directly from Equation~\eqref{eq:Idef}: at $y=y_0$ the argument of the greybody factors is $x y_0 = x/(4\pi)$. Consequently, the entire integrand depends only on the integration variable $x$ and on the dimensionless greybody functions $\Gamma_{s\ell}$ determined by the background geometry. Neither $M$ nor the EFT parameter $\epsilon$ appears explicitly in the expressions for $\mathcal{I}_0$ or $\mathcal{I}'_0$. Therefore
\begin{equation}
    \mathcal{I}_0 = \text{const}, \qquad
    \mathcal{I}_0' = \text{const}.
\end{equation}

Substituting the expansion of $\mathcal{I}(y)$ into Equation~\eqref{eq:masterflux}, and using
\begin{equation}
    T_H^2 = \frac{T_p^2}{(8\pi)^2} \left(\frac{M_p}{M}\right)^2 \left[1+4\pi c_6 \left(\frac{M_p}{M}\right)^4\right] + \mathcal{O}\left(c_6^2\left(\frac{M_p}{M}\right)^8\right),
\end{equation}
one can rewrite the evaporation rate as
\begin{equation}
\label{eq:greybodyevap}
    \frac{dM}{dt} = - \frac{\mathcal{I}_0}{128\pi^3}\frac{M_p}{t_p} \left(\frac{M_p}{M}\right)^2 \left[1 - 2\pi \rho ~ c_6 \left(\frac{M_p}{M}\right)^4 \right] + \mathcal{O}\left(c_6^2\left(\frac{M_p}{M}\right)^{10}\right),
\end{equation}
with
\begin{align}
    \rho &= \frac{3}{8\pi}\frac{\mathcal{I}_0'}{\mathcal{I}_0} - 2. 
\end{align}

Equation~\eqref{eq:greybodyevap} shows that the inclusion of greybody factors does not alter the functional mass dependence of the evaporation law. The leading term retains the standard $M^{-2}$ scaling, while the cubic curvature corrections produce a subleading contribution proportional to $M^{-6}$, exactly as found in the Stefan--Boltzmann analysis. Greybody effects modify only the numerical coefficients through $\mathcal{I}_0$ and $\rho$.

\vspace{4mm}

\noindent The Stefan--Boltzmann result may be recovered by taking the geometric--optics limit of the greybody factors. In this regime, the partial--wave sum is related to the absorption cross section via
\begin{equation}
    \sum_{\ell} (2\ell+1)\Gamma_{s\ell}(\omega) = \frac{\omega^2}{\pi}\,\sigma_{\rm abs}(\omega).
\end{equation}
For a Schwarzschild black hole the high--frequency absorption cross section approaches the geometric value \cite{Page1976a}
\begin{equation}
    \sigma_{\rm abs}(\omega) \rightarrow \sigma_{\rm geo} = \text{ const.}
\end{equation}
Substituting this into Equation~\eqref{eq:Idef} gives
\begin{align}
    \mathcal{I}(y) &= \int_0^\infty \frac{x}{e^x\pm1} \left[\frac{(xy)^2}{\pi}\sigma_{\rm geo}\right]dx \\
        &= \frac{\sigma_{\rm geo} y^2}{\pi} \int_0^\infty \frac{x^3}{e^x\pm1}dx.
\end{align}
For bosonic emission the thermal integral evaluates to
\begin{equation}
    \int_0^\infty \frac{x^3}{e^x-1}dx = \frac{\pi^4}{15}.
\end{equation}
Hence
\begin{equation}
    \mathcal{I}(y) = \frac{\sigma_{\rm geo}\pi^3}{15}\,y^2 .
\end{equation}
Differentiating $\mathcal I(y)$ yields
\begin{equation}
    \mathcal I'(y) = \frac{2\sigma_{\rm geo}\pi^3}{15}\,y ,
\end{equation}
so that
\begin{equation}
    \frac{\mathcal I_0'}{\mathcal I_0} = \frac{2}{y_0} = 8\pi .
\end{equation}

\noindent Substituting these values into Equation~\eqref{eq:greybodyevap} gives $\rho=1$, and the evaporation law reduces to
\begin{equation}
    \frac{dM}{dt} \propto -\frac{M_p}{t_p} \left(\frac{M_p}{M}\right)^2 \left[1-2\pi c_6\left(\frac{M_p}{M}\right)^4\right] + \mathcal{O}\left(c_6^2\left(\frac{M_p}{M}\right)^{10}\right), 
\end{equation}
which is precisely the Stefan--Boltzmann result obtained earlier in Equation~\eqref{eq:correctedDiffEq}.

\vspace{4mm}

Since both $\mathcal{I}_0$ and $\mathcal{I}_0'$ are mass–independent constants, $\rho$ is likewise independent of $M$ and of the expansion parameter $\epsilon$. Therefore, the critical mass can be found using the same methods as before; its existence follows from the vanishing of the leading part of Equation~\eqref{eq:greybodyevap}. Writing
\begin{equation}
\frac{dM}{dt} \propto \left(\frac{M_p}{M}\right)^2
\left[
1-2\pi \rho~ c_6\left(\frac{M_p}{M}\right)^4
\right],
\end{equation}
one finds
\begin{equation}
    M_{\rm crit, GB} = (2\pi \rho~ c_6)^{1/4} M_p = \rho^{\frac{1}{4}} M_{\rm crit, BB}.
\end{equation}
The presence of greybody factors therefore shifts the numerical value of the critical mass but does not remove the freeze--out mechanism identified in the Stefan--Boltzmann approximation.

\vspace{5mm}

\noindent Before proceeding, it is useful to comment briefly on the expected sign and magnitude of the coefficient $\rho$. The quantity $\rho$ is determined by the ratio $\mathcal I_0'/\mathcal I_0$, where $\mathcal I'(y)$ measures the response of the emission integral to a change in the dimensionless parameter $y=T_H r_H$. Increasing $y$ increases the argument $xy$ of the greybody factors. Since the transmission probabilities $\Gamma_{s\ell}(\omega r_H)$ interpolate monotonically from $0$ at low frequency to $1$ at high frequency \cite{Page1976a,Page1976b,FabbriNavarro2005}, larger values of $y$ generally correspond to larger transmission probabilities for the modes that dominate the thermal integral. Consequently, one expects $\mathcal I'(y)>0$ in the relevant region around $y_0=\frac{1}{4\pi}$.

In the geometric--optics limit discussed above this behavior yields $\mathcal I_0'/\mathcal I_0=8\pi$, giving $\rho=1$. More realistic greybody spectra suppress low--frequency modes relative to this limit but preserve the same qualitative monotonic behavior in $\mathcal{I}_0$ and $\mathcal{I}'_0$. As a result $\mathcal I_0'/\mathcal I_0$ remains positive and of comparable order to the geometric--optics limit, implying that $\rho$ is likewise expected to be positive and $\mathcal O(1)$ for Schwarzschild black holes. The presence of greybody factors therefore modifies the numerical value of the critical mass but does not alter the qualitative structure of the evaporation law.

\vspace{4mm}

\noindent A related question concerns the consistency of the result with the regime of validity of the effective field theory. The EFT expansion employed in this work assumes
\begin{equation}
    \epsilon \sim c_6\left(\frac{M_p}{M}\right)^4 \ll 1 ,
\end{equation}
so that the expansion breaks down when $M \sim c_6^{1/4} M_p$. In the Stefan--Boltzmann--blackbody approximation one finds $M_{\rm crit, BB}=c_6^{1/4}M_p$, implying that the freeze--out scale coincides parametrically with the point where the EFT ceases to be controlled.

The inclusion of greybody factors modifies the critical mass to
\begin{equation}
    M_{\rm crit,GB} = (2\pi \rho~c_6)^{1/4} M_p .
\end{equation}

A potential concern is that greybody corrections could generate a parametrically large value of 
$\rho$, thereby pushing the critical mass far above the EFT breakdown scale. This possibility can be excluded by analyzing the structure of $\mathcal{I}_0$ and $\mathcal{I}_0'$ more carefully. From Equation~\eqref{eq:Idef} one finds
\begin{equation}
    \mathcal{I}_0 = \sum_{s,\ell}(2\ell+1)\int_0^\infty \frac{x\Gamma_{s\ell}(x y_0)}{e^x\pm1}dx,
    \qquad
    \mathcal{I}_0' = \sum_{s,\ell}(2\ell+1)\int_0^\infty \frac{x^2\Gamma'_{s\ell}(x y_0)}{e^x\pm1}dx,
\end{equation}
where the derivative acts on the argument $xy$. The thermal weight $x/(e^{x}\pm 1)$ localizes both integrals to $x = \mathcal{O}(1)$, so that the greybody factors are probed only in the regime $z = \omega r_H = x y_0 = \mathcal{O}(1)$. In this region, the transmission coefficients $\Gamma(z)$ interpolate monotonically from $0$ to $1$ over a range $\Delta z \sim \mathcal{O}(1)$ and satisfies $0 \leq \Gamma_{s \ell}(z) \leq 1$. It follows that
\begin{equation}
    |\Gamma'_{s\ell}(z)| \lesssim \mathcal{O}(1)
\end{equation}
over the support of the thermal integral, so that no parametrically large gradients can develop without introducing structure on scales $\Delta z \ll 1$, which is incompatible with the smooth potential barrier. Using these properties together with the boundedness of the thermal kernel, one obtains the parametric estimates
\begin{equation}
    \mathcal{I}_0 \sim \mathcal{O}(1),
    \qquad
    \mathcal{I}_0' \sim \mathcal{O}(1),
\end{equation}
and hence
\begin{equation}
    \frac{\mathcal{I}_0'}{\mathcal{I}_0} = \mathcal{O}(1).
\end{equation}
More sharply, since both integrals receive support only from a finite interval $x = \mathcal{O}(1)$ and involve bounded integrands, their ratio is bounded by a numerical constant of order unity determined by the width of the greybody transition region. Consequently, the coefficient $\rho = \frac{3}{8\pi}\frac{\mathcal{I}_0'}{\mathcal{I}_0} - 2$ cannot become parametrically large. Achieving $\rho \gg 1$ would require greybody factors with parametrically sharp step-like behavior, $\Delta (\omega r_H) \ll 1$, which cannot arise from the smooth scattering potentials governing wave propagation in black hole spacetimes.

It follows that
\begin{equation}
    M_{\rm crit,GB} \sim c_6^{1/4} M_p ,
\end{equation}
up to an overall numerical factor of order unity. Greybody effects may therefore shift the precise value of the freeze--out mass, but they cannot move it parametrically far away from the scale where the EFT expansion $\epsilon \sim c_6(M_p/M)^4$ ceases to be small. The conclusion that the freeze--out scale and the EFT breakdown scale coincide is therefore robust.

\subsection{Sparsity of the Hawking Radiation}\label{sec:sparsity}
Another quantity that has recently received considerable attention is the sparsity of Hawking radiation \cite{AlonsoSerrano2018,Gray2016}. Rather than characterizing the total emitted power, the sparsity measures how frequently Hawking quanta are emitted relative to their characteristic oscillation time. Equivalently, it compares the mean time between successive emitted quanta with the inverse of their typical frequency. For Schwarzschild black holes this ratio is much larger than unity, indicating that Hawking radiation is an extremely dilute process in which individual quanta are emitted one at a time rather than as an approximately continuous thermal flux.

Within the Stefan--Boltzmann approximation employed throughout this work, the
number flux scales schematically as
\begin{equation}
    \dot N \sim A_H T_H^3,
\end{equation}
while the characteristic frequency of the emitted quanta satisfies $\omega_{\rm typ}\sim T_H$. Consequently, the sparsity parameter obeys the
parametric scaling
\begin{equation}
    \eta \sim \frac{\omega_{\rm typ}}{\dot N}
    \sim \frac{1}{A_H T_H^2},
\end{equation}
up to a numerical factor determined by the precise definition of the sparsity measure and by the greybody factors \cite{Gray2016,AlonsoSerrano2018}.

For a classical Schwarzschild black hole one has $A_H\propto M^2$ and $T_H\propto M^{-1}$, so that $A_H T_H^2=\mathrm{const.}$ Consequently, the sparsity remains approximately constant throughout the evaporation process. The cubic curvature corrections considered here modify both the horizon radius and the Hawking temperature, but to first order one finds
\begin{equation}
    A_H T_H^2 = \frac{\ell_p^2 T_p^2}{4\pi} \left[1 -4\pi c_6 \left(\frac{M_p}{M}\right)^4 \right] +\mathcal O\!\left(c_6^2 \left(\frac{M_p}{M}\right)^8 \right),
\end{equation}
so that the sparsity receives only perturbative corrections away from its Schwarzschild value while the EFT expansion remains under control.

The situation changes qualitatively near the characteristic mass scale $M_{\rm crit}$ for negative $c_6$, where the truncated expression for the
Hawking temperature tends to zero. Since the horizon area remains finite whereas $T_H\rightarrow 0$, the number flux tends to zero, implying
\begin{equation}
    \eta \rightarrow \infty
\end{equation}
within the leading-order approximation. Physically, this reflects that the interval between successive Hawking quanta becomes arbitrarily long, so that the radiation process effectively freezes out.

For positive $c_6$, the mechanism leading to large sparsity is different. In this case the Hawking temperature remains finite within the perturbative solution, but the corrected evaporation law develops a simple zero at the characteristic mass $M_{\rm crit}^{+}$, so that the emitted power vanishes while the typical energy of an emitted quantum remains finite. Since the particle number flux is related parametrically to the energy flux through $\dot N \sim |\dot M|/\omega_{\rm typ}$ with $\omega_{\rm typ}\sim T_H$, it follows that $\dot N\rightarrow0$ as $M\rightarrow M_{\rm crit}^{+}$. Consequently, the mean time between successive emitted quanta diverges and the sparsity parameter again satisfies
\begin{equation}
    \eta\rightarrow\infty .
\end{equation}
Thus, although the physical origin differs from the $c_6<0$ case, both signs of the cubic curvature coupling predict an increasingly sparse Hawking flux as the characteristic freeze-out scale is approached within the truncated EFT description.

As emphasized throughout this work, however, this divergence should not be interpreted as a prediction of the complete quantum-corrected theory. It occurs precisely when the expansion parameter $\epsilon\sim c_6(M_p/M)^4$ becomes of order unity, where higher-order EFT corrections neglected in the present analysis are expected to modify both the temperature and the emission spectrum. The robust conclusion is therefore not the literal divergence of the sparsity parameter, but rather the qualitative trend that the Hawking radiation becomes increasingly sparse as the evaporation approaches the characteristic freeze-out scale.

\subsection{Charged and Rotating Black Holes}\label{sec:general}
We now examine whether the conclusions derived for Schwarzschild black holes extend to more general solutions carrying electric charge and angular momentum \cite{Wald1984}. Our aim is to determine whether these additional parameters modify the parametric structure underlying the evaporation dynamics and the onset of higher-curvature effects.

A key observation is that the validity of the effective field theory expansion is controlled by local curvature invariants, rather than by global charges such as $Q$ or $J$. The relative size of higher-curvature corrections is determined by the ratio of curvature-cubed terms to the Einstein--Hilbert term, which scales parametrically as
\begin{equation}
    \epsilon \sim c_6 \frac{\mathcal{R}^2}{M_p^2}.
\end{equation}
Since curvature has dimensions of inverse length squared, one has $\mathcal{R} \sim 1/L^2$, where $L$ is the characteristic local curvature radius. This implies
\begin{equation}\label{eq:leneps}
    \epsilon \sim c_6 \left(\frac{\ell_p}{L}\right)^4.
\end{equation}

The relevant question is therefore how the curvature radius $L$ depends on the black hole parameters. For stationary black holes, the curvature near the horizon is set by the horizon radius $R_+$ up to order-one factors. For Kerr-Newman black hole solutions with charge $Q$ and spin $J$, the outer horizon radius is given by \cite{Newman1965, Poisson2004}
\begin{align}\label{eq:outrad}
    R_+ &= \ell_p \left(\frac{M}{M_p}\right) \left(1 + \sqrt{1 - J^2 \left(\frac{M_p}{M}\right)^4 - Q^2 \left(\frac{M_p}{M}\right)^2}  \right).
\end{align}
Away from extremality (i.e. when $J^2 \left(\frac{M_p}{M}\right)^4 + Q^2 \left(\frac{M_p}{M}\right)^2 \ll 1$), one generically has
\begin{equation}
    R_+ \sim \ell_p \left(\frac{M}{M_p}\right).
\end{equation}
It follows that the curvature scale near the horizon is set by $L \sim R_+ \sim \ell_p \left(\frac{M}{M_p}\right)$, independently of the precise values of $Q$ and $J$, as long as the black hole is not parametrically close to extremality.

Substituting this scaling into the expansion parameter \eqref{eq:leneps} yields
\begin{equation}
    \epsilon \sim c_6 \left(\frac{M_p}{M}\right)^4.
\end{equation}
The onset of higher-curvature effects is therefore governed by the same characteristic mass scale $M_{\rm crit} \sim c_6^{1/4} M_p$ as in the Schwarzschild case.

The evaporation dynamics further supports this conclusion. For charged black holes, Hawking radiation predominantly emits neutral quanta, while Schwinger pair production efficiently discharges the black hole at low temperatures \cite{Gibbons1975,HiscockWeems1990}. For rotating black holes, angular momentum is rapidly radiated away through Hawking emission and superradiance \cite{Page1976b}. In both cases, the evolution generically drives the system away from extremality, so that for most of the evaporation history one has $Q \ll \left(\frac{M}{M_p}\right)$ and $J \ll \left(\frac{M}{M_p}\right)^2$. As a result, $R_+ \sim \ell_p \left(\frac{M}{M_p}\right)$ throughout most of the evolution, and both the curvature scale and the temperature are effectively controlled by the mass alone. The parametric form of the evaporation rate therefore remains unchanged, with a leading $M^{-2}$ scaling and a subleading cubic-curvature correction proportional to $M^{-6}$.

A qualitatively different situation can arise for finely tuned configurations that remain close to extremality. From the expression for the outer horizon radius \eqref{eq:outrad} it follows that near extremality the square root becomes parametrically small. In this regime, the Hawking temperature is correspondingly suppressed, since for charged and rotating black holes \cite{Hawking1975, FrolovNovikov1998}
\begin{equation}
    T_H = \frac{T_p}{4\pi} \left(\frac{M_p}{M}\right) \frac{\sqrt{1 - J^2 \left(\frac{M_p}{M}\right)^4 - Q^2 \left(\frac{M_p}{M} \right)^2}}{1 - \frac{1}{2}Q^2 \left(\frac{M_p}{M}\right)^2+ \sqrt{1 - J^2 \left(\frac{M_p}{M}\right)^4 - Q^2 \left(\frac{M_p}{M} \right)^2}},
\end{equation}
and thus $T_H$ tends to zero as the extremal bound is approached. As a result, the Hawking temperature is strongly suppressed independently of higher-curvature corrections.

If a black hole were to remain close to extremality throughout its evolution, the evaporation dynamics would therefore be dominated by this temperature suppression rather than by the cubic curvature corrections considered in this work. In such a scenario, the approach to extremality itself induces a slow-down, and the simple scaling $\epsilon \sim c_6 (M_p/M)^4$ no longer provides a complete characterization of the late-time dynamics.

\vspace{4mm}

We conclude that the central result of this work is robust for generic evaporation: cubic curvature corrections remain perturbative until $M \sim M_{\rm crit}$, where the curvature approaches the Planck scale and the effective field theory breaks down. This behavior is generic for four-dimensional black holes and does not rely on the assumption of vanishing charge or angular momentum, but it can be modified in finely tuned near-extremal configurations.

\section{Comparison with Other Quantum Gravity Approaches}\label{sec:comparison}
A number of quantum–gravity motivated approaches predict qualitative deviations from the semiclassical evaporation law at small black hole masses, including the appearance of a maximal Hawking temperature, late–time cooling, or minimal–mass configurations. While these features are often interpreted as indications of evaporation freeze–out or remnant formation, their physical origin and theoretical status differ significantly across frameworks.

A common structural pattern is that the semiclassical evaporation law drives the system toward a regime in which additional quantum–gravitational effects become important. In many cases, the departure from standard Hawking scaling occurs at a characteristic mass scale associated with Planckian physics. However, the mechanism responsible for this behavior, and its interpretation, depend on how quantum gravity is incorporated.

In asymptotic safety, modifications arise from the renormalization–group running of gravitational couplings \cite{Reuter1998,ReuterSaueressig2012}. Renormalization–group improved black hole metrics replace the classical Newton constant by a scale–dependent quantity, leading to a maximal temperature and vanishing temperature at finite mass \cite{BonannoReuter2000,BonannoReuter2006}. In this case, the characteristic mass scale is determined by the approach to a nontrivial ultraviolet fixed point, and the modified evaporation behavior reflects the underlying renormalization–group dynamics.

In Lovelock or Einstein–Gauss–Bonnet theories \cite{BoulwareDeser1985,Cai2002}, higher–curvature terms are treated nonperturbatively as part of the fundamental action. Modified thermodynamic behavior, including extremal or minimal–mass configurations, arises from the structure of the full theory and depends sensitively on the choice of couplings. The resulting freeze–out is therefore a property of a self–consistent modified gravitational dynamics rather than a perturbative correction.

In approaches based on generalized uncertainty principles or nonlocal gravity \cite{Adler2001,Scardigli1999,Modesto2010,Biswas2012}, the evaporation law is altered through modified kinematics, minimal length effects, or nonlocal form factors. In these cases, the suppression of evaporation at small masses is built into the underlying framework and typically leads to remnant–like behavior.

The present analysis differs conceptually from these approaches in that it does not introduce new nonperturbative dynamics, running couplings, or modified kinematics. Instead, we work within a local four–dimensional gravitational effective field theory and treat higher–curvature operators perturbatively. The relevant expansion parameter is
\begin{equation}
    \epsilon \sim c_6 \left(\frac{M_p}{M}\right)^4.
\end{equation}
We find that the apparent freeze–out of evaporation occurs precisely when $\epsilon \sim 1$, at which point the derivative expansion ceases to be hierarchical.

This identifies a direct link between evaporation dynamics and the regime of validity of the effective theory. The same condition implies that the cubic correction becomes comparable to the Einstein–Hilbert term, curvature invariants at the horizon become Planckian, and higher–order operators are no longer parametrically suppressed. The characteristic freeze–out mass therefore does not lie within a controlled perturbative regime, but instead marks the breakdown of the effective field theory description.

In this sense, the qualitative similarity of late–stage evaporation across different approaches reflects a common structural feature, namely the approach to Planckian curvature. The interpretation, however, is different: in the EFT framework, the apparent freeze–out does not represent a physical endpoint of evaporation, but rather a dynamical signal that the perturbative expansion has reached the boundary of its validity. The ultimate fate of the black hole must therefore be determined by the underlying ultraviolet completion beyond the EFT regime.

\section{Discussion}

In this work we have investigated how cubic curvature corrections within a local third-order gravitational effective field theory modify the late-stage evaporation of Schwarzschild black holes. Using the perturbatively corrected horizon area and temperature derived in \cite{Calmet2021Entropy}, we obtained a modified evaporation law that departs qualitatively from the classical Hawking prediction as the black hole mass approaches the Planck scale.

The central purpose of this analysis has been to identify structural features of evaporation dynamics within a controlled truncation of gravitational effective field theory, and \textit{not} to claim the existence of physical Planck-scale remnants. In this sense, the results should be interpreted diagnostically: rather than predicting a modified endpoint of evaporation, they show that the evaporation dynamics itself becomes sensitive to the breakdown of the derivative expansion as the Planck scale is approached.

Our conclusions should therefore be understood in the standard effective-field-theory sense. The cubic curvature operator represents only the leading correction capable of modifying the Schwarzschild geometry in vacuum, and the onset of evaporation slow-down indicates that higher-order operators can no longer be neglected. The present analysis therefore identifies the boundary of validity of the truncated EFT rather than establishing the properties of the exact ultraviolet completion.

A central result is that the apparent freeze–out of evaporation arises precisely when the dimensionless expansion parameter $\epsilon \sim c_6 (M_p/M)^4$ becomes of order unity. This establishes a direct link between evaporation dynamics and the regime of validity of the effective field theory, showing that late–stage evolution provides a dynamical probe of the breakdown of perturbative control.

A characteristic mass scale,
\begin{align}
    M_{\rm crit} = (2\pi |c_6|)^{1/4} M_p,
\end{align}
emerges naturally from the perturbative expansion. As shown in Section~\ref{sec:evaporation}, this scale arises when the cubic correction becomes comparable in magnitude to the leading semiclassical contribution. Depending on the sign of the Wilson coefficient $c_6$, two distinct freeze-out mechanisms appear within the truncated theory.

For $c_6 < 0$, the first-order correction to the Hawking temperature leads to a non-monotonic temperature profile: $T_H$ reaches a maximum at $M_T^{-}$ and subsequently decreases toward zero at $M_{\rm crit}$. This behavior is reflected dynamically in the sign change of $\frac{dT_H}{dt}$, indicating a transition from heating to cooling during the evaporation process. Within the truncated expansion this manifests as a temperature-based freeze-out. However, since the vanishing occurs precisely when the expansion parameter $\epsilon \sim c_6 (M_p/M)^4$ becomes order unity, this behavior must be interpreted as a manifestation of the breakdown of perturbative control, since it occurs precisely at $\epsilon \sim 1$ where the derivative expansion ceases to be valid. 

For $c_6 > 0$, the Hawking temperature remains finite at $M_{\rm crit}$, but the leading-order mass-loss rate $\frac{dM}{dt}$ vanishes due to a cancellation between the classical Hawking term and the cubic EFT correction. This demonstrates explicitly that, once perturbative corrections are included, the conditions $T_H=0$ and $\frac{dM}{dt}=0$ are no longer equivalent. The freeze-out in this branch arises from the algebraic structure of the truncated expansion rather than from the thermodynamic vanishing of temperature. This cancellation occurs precisely at the boundary of validity of the EFT expansion, and higher–order operators are generically expected to modify or remove it once the perturbative hierarchy is lost. 

A complementary perspective is provided by the time-dependent evolution of the Hawking temperature. As shown in Section~\ref{sec:thermo}, the quantity $\frac{dT_H}{dt} = \frac{dT_H}{dM}\frac{dM}{dt}$ remains strictly positive for $c_6>0$, so that the black hole continues to heat up monotonically throughout the evaporation process. For $c_6<0$, however, $\frac{dT_H}{dt}$ develops a zero and subsequently becomes negative, implying the existence of a maximal temperature followed by a cooling phase. This non-monotonic behavior is consistent with the structure of the corrected temperature $T_H(M)$, and provides a dynamical characterization of the temperature-based freeze-out identified in the truncated expansion.

In the two branches the late-stage dynamics differ qualitatively. For $c_6>0$, the evaporation time diverges logarithmically as $M \to M_{\rm crit}$ due to the appearance of a simple zero in the mass-loss rate, implying an asymptotic approach to the critical mass. In contrast, for $c_6<0$ the mass-loss rate remains finite, and the black hole reaches $M \sim M_{\rm crit}$ in a finite time within the truncated dynamics. 

The thermodynamic analysis provides additional insight. The heat capacity $C\equiv \frac{dM}{dT_H}$ exhibits sign-dependent behavior that mirrors the temperature-based and cancellation-based freeze-out mechanisms. For $c_6<0$, the heat capacity becomes positive below $M_T$, enabling cooling toward $T_H \to 0$ within the truncated description. For $c_6>0$, the heat capacity remains negative even as the mass-loss rate vanishes, reflecting the non-thermodynamic origin of freeze-out in that branch. 

The thermodynamic interpretation is further clarified by the corrected first law of black hole thermodynamics, $dM=T_H dS$, which continues to hold to first order in the EFT expansion when the corrected Wald entropy is used \cite{Calmet2021Entropy}. In particular, the change of sign of the heat capacity for $c_6<0$ occurs smoothly: $C$ remains continuous, passes through zero at $M_T^{-}$, and becomes positive without developing a divergence. Consequently, the truncated EFT does not predict a genuine second-order thermodynamic phase transition, but rather a continuous crossover from the familiar Schwarzschild-like regime with negative heat capacity to a low-mass regime in which the black hole cools as it loses mass. Furthermore, although the Hawking temperature tends to zero at $M_{\rm crit}^{-}$ within the truncated approximation, the heat capacity remains finite, consistent with the first law through the corresponding behavior of the entropy.

A key result of this work is that the characteristic freeze–out mass coincides parametrically with the onset of Planckian curvature at the horizon. As shown in Section~\ref{sec:kretschmann}, the Kretschmann invariant evaluated at $M_{\rm crit}$ is of order $1/\ell_p^4$ for $|c_6|=\mathcal{O}(1)$. This demonstrates that the scale at which evaporation slows down is not parametrically separated from the scale at which the curvature expansion breaks down. The freeze–out therefore does not occur within a controlled regime of the effective theory, but instead marks the loss of its validity.

The parametric comparison with quartic curvature operators further strengthens this conclusion. At $M \sim M_{\rm crit}$, generic quartic invariants contribute at the same order as the cubic term unless their Wilson coefficients are tuned to be parametrically small. The derivative expansion therefore ceases to be hierarchical exactly where the freeze-out behavior emerges. This shows that the cubic truncation alone cannot determine the ultimate endpoint of evaporation; rather, it provides a dynamical indication that the truncated description has reached the boundary of its applicability and that the full tower of higher–curvature operators must be included. 

The extensions discussed in Section~\ref{sec:extensions} further clarify the structural robustness of these conclusions. In particular, the explicit greybody analysis of Section~\ref{sec:greybody} shows that the full emission rate can be written in the form $\frac{dM}{dt} \propto T_H^2\,\mathcal{I}(T_H r_H)$, where the function $\mathcal{I}$ depends only on the combination $y=T_H r_H$. Expanding around the Schwarzschild value $y_0=\frac{1}{4\pi}$ demonstrates that greybody factors modify only the numerical coefficients of the evaporation law, while the mass dependence and the $M^{-6}$ cubic-curvature correction remain unchanged.

Charged and rotating black holes introduce additional parameters, but these do not alter the parametric structure that controls the effective field theory expansion. As discussed in Section~\ref{sec:general}, the expansion parameter is determined by local curvature invariants rather than by global charges such as $Q$ or $J$. Away from extremality, the horizon radius scales as $R_+ \sim \ell_p \left(M/M_p\right)$, so that the curvature scale is set by the mass and the expansion parameter retains the universal form $\epsilon \sim c_6 (M_p/M)^4$.

The evaporation dynamics further drive generic solutions away from extremality, through discharge in the charged case and spin-down in the rotating case. As a result, the late-time evolution is well approximated by the Schwarzschild analysis, and the onset of higher-curvature effects remains tied to $M \sim M_{\rm crit}$. Only finely tuned near-extremal configurations can deviate from this behavior, but such configurations are not expected to arise dynamically. The conclusion that evaporation slow-down coincides with EFT breakdown and Planckian curvature is therefore robust and does not rely on the absence of charge or angular momentum.

The relation to other quantum gravity approaches can now be understood more precisely. While a variety of frameworks exhibit qualitatively similar late–stage behavior (such as maximal temperatures or apparent minimal masses), the physical origin of these effects differs. In asymptotic safety, the modification is tied to the renormalization–group flow of gravitational couplings, while in Lovelock-type theories it follows from treating higher–curvature terms nonperturbatively as part of the fundamental action. In GUP and nonlocal models, the evaporation law is altered through modified kinematics or nonlocal dynamics.

In contrast, the present analysis does not introduce new dynamical ingredients beyond the local effective field theory expansion. Instead, the condition
\begin{equation}
    c_6 \left(\frac{M_p}{M}\right)^4 \sim 1
\end{equation}
simultaneously marks the onset of Planckian curvature, the breakdown of the perturbative hierarchy, and the appearance of evaporation slow–down. The qualitative similarity of freeze–out behavior across different approaches therefore reflects a common structural feature (the approach to Planckian curvature), while in the EFT framework it acquires a distinct interpretation as a dynamical signal of the breakdown of perturbative control rather than a prediction of a physical endpoint. The main conclusion is therefore that late–stage black hole evaporation provides a dynamical probe of the boundary of validity of gravitational effective field theory.

In summary, cubic curvature corrections in gravitational effective field theory qualitatively modify late-stage black hole evaporation, producing evaporation slow–down and apparent freeze–out precisely at the scale where the derivative expansion breaks down. The analysis shows that late–stage evaporation dynamics encode the boundary of validity of the effective theory, providing a dynamical criterion for its breakdown rather than a purely kinematical estimate. While the existence of a physical Planck-scale remnant cannot be established within this truncated framework, the results demonstrate that local higher–curvature operators already drive the system to the edge of perturbative control. Determining the true endpoint of evaporation therefore requires either inclusion of higher-order operators beyond cubic order or input from a complete ultraviolet theory of quantum gravity.

\subsection{Outlook}
The results of this work suggest that late–stage black hole evaporation provides a sensitive probe of the regime of validity of gravitational effective field theory. In particular, the identification of evaporation slow–down with the condition $\epsilon \sim c_6 (M_p/M)^4 \sim 1$ indicates that dynamical observables can be used to identify the breakdown of the derivative expansion as Planckian curvature is approached.

A natural extension of this analysis is to go beyond the cubic truncation by systematically including higher–order curvature operators. Since quartic and higher invariants are expected to contribute at the same order near $M \sim M_{\rm crit}$, such an extension would clarify whether the qualitative freeze–out behavior persists, is shifted, or is removed once the effective theory is treated more completely.

Another important direction is the explicit construction of corrected solutions and thermodynamics for charged and rotating black holes within the same EFT framework. This would allow for a more detailed analysis of the interplay between higher–curvature effects and near–extremal dynamics, and would test the robustness of the connection between evaporation slow–down and the onset of Planckian curvature.

It would also be of interest to further relate the EFT description to other quantum gravity approaches. While the present work interprets late–stage modifications as a signal of perturbative breakdown, other frameworks attribute similar behavior to nonperturbative dynamics, running couplings, or modified kinematics. Establishing more direct connections between these perspectives may help to disentangle universal features of black hole evaporation from model–dependent effects.

Finally, a complete understanding of the endpoint of evaporation requires input from the ultraviolet completion of gravity. The EFT analysis presented here provides a controlled description up to the boundary of its validity, but determining the fate of the system beyond this regime remains an open problem.

\section{Acknowledgments}
\noindent We would like to thank Xavier Calmet and Folkert Kuipers for their valuable insights and helpful correspondence regarding the results presented in this work.

\appendix

\bibliography{CQG_Submission}

@article{Hawking1974Explosions,
  title = {{Black Hole Explosions}},
  year = {1974},
  journal = {Nature (London)},
  author = {Hawking, S. W.},
  volume = {248},
  url = {https://doi.org/10.1038/248030a0}
}

@article{MacGibbon1987Close,
  title = {{Can Planck-mass Relics of Evaporating Black Holes Close the Universe?}},
  year = {1987},
  journal = {Nature (London)},
  author = {MacGibbon, J. H.},
  volume = {329},
  url = {https://doi.org/10.1038/329308a0}
}

@article{Bowick1988Axionic,
  title = {{Axionic Black Holes and an Aharonov-Bohm Effect for Strings}},
  year = {1988},
  journal = {Phys. Rev. Lett.},
  author = {Bowick, M. J. and Giddings, S. B. and Harvey, J. A. and Horowitz, G. T. and Strominger, A.},
  volume = {61},
  url = {https://doi.org/10.1103/PhysRevLett.61.2823}
}

@article{Calmet2021Entropy,
  title = {{Quantum Gravitational Corrections to the Entropy of a Schwarzschild Black Hole}},
  year = {2021},
  journal = {Phys. Rev. D},
  author = {Calmet, X. and Kuipers, F.},
  volume = {104},
  url = {https://doi.org/10.1103/PhysRevD.104.066012}
}

@article{Barrow1992Cosmology,
  title = {{The Cosmology of Black Hole Relics}},
  year = {1992},
  journal = {Phys. Rev. D},
  author = {Barrow, J. D. and Copeland, E. J. and Liddle, A. R.},
  volume = {46},
  url = {https://doi.org/10.1103/PhysRevD.46.645}
}

@article{Aharonov1987Unitarity,
  title = {{The Unitarity Puzzle and Planck Mass Stable Particles}},
  year = {1987},
  journal = {Phys. Lett. B},
  author = {Aharonov, Y. and Casher, A. and Nussinov, S.},
  volume = {191},
  url = {https://doi.org/10.1016/0370-2693(87)91320-7}
}

@article{Banks1992Horned,
  title = {{Are Horned Particles the end Point of Hawking Evaporation?}},
  year = {1992},
  journal = {Phys. Rev. D},
  author = {Banks, T. and Dabholkar, A. and Douglas, M. R. and O'Loughlin, M.},
  volume = {45},
  url = {https://doi.org/10.1103/PhysRevD.45.3607}
}

@article{Banks1993Information,
  title = {{Black Hole Remnants and the Information Puzzle}},
  year = {1993},
  journal = {Phys. Rev. D},
  author = {Banks, T. and O'Loughlin, M. and Strominger, A.},
  volume = {47},
  url = {https://doi.org/10.1103/PhysRevD.47.4476}
}

@article{Callan1989String,
  title = {{Black Holes in String Theory}},
  year = {1989},
  journal = {Nucl. Phys. B},
  author = {Callan, C. G. and Myers, R. C. and Perry, M. J.},
  volume = {311},
  url = {https://doi.org/10.1016/0550-3213(89)90172-7}
}

@article{Myers1988Lovelock,
  title = {{Black-Hole Thermodynamics in Lovelock Gravity}},
  year = {1988},
  journal = {Phys. Rev. D},
  author = {Myers, R. C. and Simon, J. Z.},
  volume = {38},
  url = {https://doi.org/10.1103/PhysRevD.38.2434}
}

@article{Whitt1988Spherically,
  title = {{Spherically Symmetric Solutions of General Second-order Gravity}},
  year = {1988},
  journal = {Phys. Rev. D},
  author = {Whitt, B.},
  volume = {38},
  url = {https://doi.org/10.1103/PhysRevD.38.3000}
}

@article{Alexeyev2002String,
  title = {{Black-hole Relics in String Gravity: Last Stages of Hawking Evaporation}},
  year = {2002},
  journal = {Class. Quantum Grav.},
  author = {Alexeyev, S. and Barrau, A. and Boudoul, G. and Khovanskaya, O. and Sazhin, M.},
  volume = {19},
  url = {http://dx.doi.org/10.1088/0264-9381/19/16/314}
}

@article{Giddings1992Massive,
  title = {{Black Holes and Massive Remnants}},
  year = {1992},
  journal = {Phys. Rev. D},
  author = {Giddings, B. S.},
  volume = {46},
  url = {http://dx.doi.org/10.1103/PhysRevD.46.1347}
}

@article{Chen2015Information,
  title = {{Black Hole Remnants and the Information Loss Paradox}},
  year = {2015},
  journal = {Physics Reports},
  author = {Chen, P. and Ong, Y. C. and Yeom, D. H.},
  volume = {603},
  url = {https://doi.org/10.1016/j.physrep.2015.10.007}
}

@article{Barrau2019DarkMatter,
  title = {{Dark Matter as Planck Relics Without Too Exotic Hypotheses}},
  year = {2019},
  journal = {Phys. Rev. D},
  author = {Barrau, A. and Martineau, K. and Moulin, F. and Ngono, J. F.},
  volume = {100},
  url = {https://doi.org/10.1103/PhysRevD.100.123505}
}

@article{Calmet2018VanishingCurvature,
  title = {{Vanishing of quantum gravitational corrections to vacuum solutions of general relativity at second order in curvature}},
  year = {2018},
  journal = {Phys. Lett. B},
  author = {Calmet, X.},
  month = {12},
  pages = {36--38},
  volume = {787},
  publisher = {Elsevier B.V.},
  issn = {03702693},
  url = {https://doi.org/10.1016/j.physletb.2018.10.040}
}

@article{Calmet2018Quantum,
  title = {{Quantum Corrections to Schwarzschild Black Hole}},
  year = {2018},
  journal = {The Eur. Phys. J. C},
  author = {Calmet, X. and El-Menoufi, B. K.},
  volume = {77},
  url = {https://doi.org/10.1140/epjc/s10052-017-4802-0}
}

@article{Donoghue2012Effective,
  title = {{The Effective Field Theory Treatment of Quantum Gravity}},
  year = {2012},
  journal = {AIP Conference Proceedings},
  author = {Donoghue, J. F.},
  volume = {1483},
  url = {https://doi.org/10.1063/1.4756964}
}

@article{Burgess2004Everyday,
  title = {{Quantum Gravity in Everyday Life: General Relativity as an Effective Field Theory}},
  year = {2004},
  journal = {Living Rev. Relativity},
  author = {Burgess, C. P.},
  volume = {7},
  url = {https://doi.org/10.12942/lrr-2004-5}
}

@article{Saraswat2017Weak,
  title = {{Weak Gravity Conjecture and Effective Field Theory}},
  year = {2017},
  journal = {Phys. Rev. D},
  author = {Saraswat, P.},
  volume = {95},
  url = {https://doi.org/10.1103/PhysRevD.95.025013}
}

@article{Ruhdorfer2020Effective,
  title = {{Effective Field Theory of Gravity to all Orders}},
  year = {2020},
  journal = {J. High Energy Phys.},
  author = {Ruhdorfer, M. and Serra, J. and Weiler, A.},
  volume = {83},
  url = {https://doi.org/10.1007/JHEP05(2020)083}
}

@article{Goroff1986Ultraviolet,
  title = {{The Ultraviolet Behavior of Einstein Gravity}},
  year = {1986},
  journal = {Nucl. Phys. B},
  author = {Goroff, M. H. and Sagnotti, A.},
  volume = {266},
  url = {https://doi.org/10.1016/0550-3213(86)90193-8}
}

@article{Donoghue1994Leading,
  title = {{General Relativity as an effective field theory: The leading quantum corrections}},
  year = {1994},
  journal = {Phys. Rev. D},
  author = {Donoghue, J. F.},
  volume = {50},
  url = {https://doi.org/10.1103/PhysRevD.50.3874}
}

@article{Calmet2017QuantumCorrections,
  title = {{Quantum corrections to Schwarzschild black hole}},
  year = {2017},
  journal = {The Eur. Phys. J. C},
  author = {Calmet, X. and El-Menoufi, B. K.},
  volume = {77},
  url = {https://doi.org/10.1140/epjc/s10052-017-4802-0}
}

@article{Reuter1998,
  title = {{Nonperturbative evolution equation for quantum gravity}},
  year = {1998},
  journal = {Phys. Rev. D},
  author = {Reuter, M.},
  volume = {57},
  url = {https://doi.org/10.1103/PhysRevD.57.971}
}

@article{ReuterSaueressig2012,
  title = {{Quantum Einstein Gravity}},
  year = {2012},
  journal = {New J. Phys.},
  author = {Reuter, M. and Saueressig, M.},
  volume = {14},
  url = {https://doi.org/10.1088/1367-2630/14/5/055022}
}

@article{BonannoReuter2000,
  title = {{Renormalization group improved black hole spacetimes}},
  year = {2000},
  journal = {Phys. Rev. D},
  author = {Bonanno, A. and Reuter, M.},
  volume = {62},
  url = {https://doi.org/10.1103/PhysRevD.62.043008}
}

@article{BonannoReuter2006,
  title = {{Spacetime structure of an evaporating black hole in quantum gravity}},
  year = {2006},
  journal = {Phys. Rev. D},
  author = {Bonanno, A. and Reuter, M.},
  volume = {73},
  url = {https://doi.org/10.1103/PhysRevD.73.083005}
}

@article{BoulwareDeser1985,
  title = {{String-generated gravity models}},
  year = {1985},
  journal = {Phys. Rev. Lett.},
  author = {Boulware, D. G. and Deser, S.},
  volume = {55},
  url = {https://doi.org/10.1103/PhysRevLett.55.2656}
}

@article{Cai2002,
  title = {{Gauss-Bonnet black holes in AdS spaces}},
  year = {2002},
  journal = {Phys. Rev. D},
  author = {Cai, R.},
  volume = {65},
  url = {https://doi.org/10.1103/PhysRevD.65.084014}
}

@article{Adler2001,
  title = {{The Generalized Uncertainty Principle and Black Hole Remnants}},
  year = {2001},
  journal = {Gen. Rel. Grav.},
  author = {Adler, J. A. and Chen, P. and Santiago, D. I.},
  volume = {33},
  url = {https://doi.org/10.1023/A:1015281430411}
}

@article{Scardigli1999,
  title = {{Generalized uncertainty principle in quantum gravity from micro-black hole gedanken experiment}},
  year = {1999},
  journal = {Phys. Lett. B},
  author = {Scardigli, F.},
  volume = {452},
  url = {https://doi.org/10.1016/S0370-2693(99)00167-7}
}

@article{Modesto2010,
  title = {{Model for nonsingular black hole collapse and evaporation}},
  year = {2010},
  journal = {Phys. Rev. D},
  author = {Hossenfelder, S. and Modesto, L. and Prémont-Schwarz, I.},
  volume = {81},
  url = {https://doi.org/10.1103/PhysRevD.81.044036}
}

@article{Biswas2012,
  title = {{Towards Singularity- and Ghost-Free Theories of Gravity}},
  year = {2012},
  journal = {Phys. Rev. D},
  author = {Biswas, T. and Gerwick, E. and Koivisto, T. and Mazumdar, A.},
  volume = {108},
  url = {https://doi.org/10.1103/PhysRevLett.108.031101}
}

@article{Page1976a,
  title = {{Particle emission rates from a black hole: Massless particles from an uncharged, nonrotating hole}},
  author = {Page, Don N.},
  journal = {Phys. Rev. D},
  volume = {13},
  number = {2},
  pages = {198--206},
  year = {1976},
  doi = {10.1103/PhysRevD.13.198},
  url = {https://doi.org/10.1103/PhysRevD.13.198}
}

@article{Page1976b,
  title = {{Particle emission rates from a black hole. II. Massless particles from a rotating hole}},
  author = {Page, Don N.},
  journal = {Phys. Rev. D},
  volume = {14},
  number = {12},
  pages = {3260--3273},
  year = {1976},
  doi = {10.1103/PhysRevD.14.3260},
  url = {https://doi.org/10.1103/PhysRevD.14.3260}
}

@book{FabbriNavarro2005,
  title = {{Modeling Black Hole Evaporation}},
  author = {Fabbri, Alessandro and Navarro-Salas, Jos\'e},
  publisher = {Imperial College Press / World Scientific},
  year = {2005},
  address = {London, Singapore},
  isbn = {9781860945274},
  url = {https://www.worldscientific.com/worldscibooks/10.1142/6117}
}

@article{VandeVen1992,
  title = {{Two-loop quantum gravity}},
  author = {van de Ven, A. E. M.},
  journal = {Nucl. Phys. B},
  volume = {378},
  pages = {309--366},
  year = {1992},
  doi = {10.1016/0550-3213(92)90011-Y},
  url = {https://doi.org/10.1016/0550-3213(92)90011-Y}
}

@book{Wald1984,
  title = {{General Relativity}},
  author = {Wald, Robert M.},
  publisher = {University of Chicago Press},
  year = {1984}
}

@book{Carroll2004,
  title = {{Spacetime and Geometry: An Introduction to General Relativity}},
  author = {Carroll, Sean M.},
  publisher = {Addison-Wesley},
  year = {2004}
}

@article{BuenoCano2016,
  title = {{Four-dimensional black holes in Einsteinian cubic gravity}},
  author = {Bueno, Pablo and Cano, Pablo A.},
  journal = {Phys. Rev. D},
  volume = {94},
  pages = {104005},
  year = {2016},
  doi = {10.1103/PhysRevD.94.104005},
  url = {https://doi.org/10.1103/PhysRevD.94.104005}
}

@article{Hennigar2017,
  title = {{Black holes in Einsteinian cubic gravity}},
  author = {Hennigar, Robie A. and Mann, Robert B.},
  journal = {Phys. Rev. D},
  volume = {95},
  pages = {064055},
  year = {2017},
  doi = {10.1103/PhysRevD.95.064055},
  url = {https://doi.org/10.1103/PhysRevD.95.064055}
}

@article{BuenoCano2017,
  title = {{Universal black hole stability in four dimensions}},
  author = {Bueno, Pablo and Cano, Pablo A.},
  journal = {Phys. Rev. D},
  volume = {96},
  pages = {024034},
  year = {2017},
  doi = {10.1103/PhysRevD.96.024034},
  url = {https://doi.org/10.1103/PhysRevD.96.024034}
}

@article{Bonanno2000,
  title = {{Renormalization group improved black hole spacetimes}},
  author = {Bonanno, Alfio and Reuter, Martin},
  journal = {Phys. Rev. D},
  volume = {62},
  pages = {043008},
  year = {2000},
  doi = {10.1103/PhysRevD.62.043008},
  url = {https://doi.org/10.1103/PhysRevD.62.043008}
}

@article{Bonanno2006,
  title = {{Spacetime structure of an evaporating black hole in quantum gravity}},
  author = {Bonanno, Alfio and Reuter, Martin},
  journal = {Phys. Rev. D},
  volume = {73},
  pages = {083005},
  year = {2006},
  doi = {10.1103/PhysRevD.73.083005},
  url = {https://doi.org/10.1103/PhysRevD.73.083005}
}

@article{DeFelice2023,
  title = {{Instabilities of Einsteinian cubic gravity}},
  author = {De Felice, Antonio and Tsujikawa, Shinji},
  journal = {Phys. Rev. D},
  volume = {107},
  pages = {064060},
  year = {2023},
  doi = {10.1103/PhysRevD.107.064060},
  url = {https://doi.org/10.1103/PhysRevD.107.064060}
}

@article{Gonzalez2023,
  title = {{On the physical viability of black holes in Einsteinian cubic gravity}},
  author = {González, Pedro A. et al.},
  journal = {Phys. Lett. B},
  volume = {844},
  pages = {138099},
  year = {2023},
  doi = {10.1016/j.physletb.2023.138099},
  url = {https://doi.org/10.1016/j.physletb.2023.138099}
}

@article{Hawking1975,
  title = {{Particle Creation by Black Holes}},
  author = {Hawking, Stephen W.},
  journal = {Commun. Math. Phys.},
  volume = {43},
  pages = {199--220},
  year = {1975},
  doi = {10.1007/BF02345020},
  url = {https://doi.org/10.1007/BF02345020}
}

@article{Stelle1977Renormalization,
  author = {Stelle, K. S.},
  title = {Renormalization of Higher Derivative Quantum Gravity},
  journal = {Phys. Rev. D},
  volume = {16},
  pages = {953--969},
  year = {1977},
  doi = {10.1103/PhysRevD.16.953}
}

@article{MyersSimon1988GB,
  author = {Myers, R. C. and Simon, J. Z.},
  title = {Black-hole thermodynamics in Lovelock gravity},
  journal = {Phys. Rev. D},
  volume = {38},
  pages = {2434--2444},
  year = {1988},
  doi = {10.1103/PhysRevD.38.2434}
}

@article{Wheeler1986Lovelock,
  author = {Wheeler, J. T.},
  title = {Symmetric Solutions to the Gauss-Bonnet Extended Einstein Equations},
  journal = {Nucl. Phys. B},
  volume = {268},
  pages = {737--746},
  year = {1986},
  doi = {10.1016/0550-3213(86)90268-3}
}

@article{PadmanabhanKothawala2013,
  author = {Padmanabhan, T. and Kothawala, D.},
  title = {Lanczos-Lovelock models of gravity},
  journal = {Phys. Rept.},
  volume = {531},
  pages = {115--171},
  year = {2013},
  doi = {10.1016/j.physrep.2013.07.002}
}

@article{Cai2004GBThermo,
  author = {Cai, Rong-Gen},
  title = {A note on thermodynamics of black holes in Lovelock gravity},
  journal = {Phys. Lett. B},
  volume = {582},
  pages = {237--242},
  year = {2004},
  doi = {10.1016/j.physletb.2003.12.015}
}

@article{CveticNojiriOdintsov2002,
  author = {Cvetič, M. and Nojiri, S. and Odintsov, S. D.},
  title = {Black hole thermodynamics and negative entropy in deSitter and anti-deSitter Einstein–Gauss–Bonnet gravity},
  journal = {Nucl. Phys. B},
  volume = {628},
  pages = {295--330},
  year = {2002},
  doi = {10.1016/S0550-3213(02)00075-5}
}

@article{LuPerkinsPopeStelle2015,
  author = {Lu, H. and Perkins, A. and Pope, C. N. and Stelle, K. S.},
  title = {Black Holes in Higher-Derivative Gravity},
  journal = {Phys. Rev. Lett.},
  volume = {114},
  pages = {171601},
  year = {2015},
  doi = {10.1103/PhysRevLett.114.171601}
}

@article{Gibbons1975,
  author = {Gibbons, G. W.},
  title = {Vacuum Polarization and the Spontaneous Loss of Charge by Black Holes},
  journal = {Communications in Mathematical Physics},
  volume = {44},
  pages = {245--264},
  year = {1975},
  doi = {10.1007/BF01609829}
}

@article{HiscockWeems1990,
  author = {Hiscock, William A. and Weems, Larissa D.},
  title = {Evolution of Charged Evaporating Black Holes},
  journal = {Physical Review D},
  volume = {41},
  pages = {1142--1151},
  year = {1990},
  doi = {10.1103/PhysRevD.41.1142}
}

@article{Newman1965,
  title = {Metric of a Rotating, Charged Mass},
  author = {Newman, E. T. and Couch, E. and Chinnapared, K. and Exton, A. and Prakash, A. and Torrence, R.},
  journal = {J. Math. Phys.},
  volume = {6},
  pages = {918--919},
  year = {1965},
  doi = {10.1063/1.1704351},
  url = {https://doi.org/10.1063/1.1704351}
}

@book{Poisson2004,
  title = {A Relativist's Toolkit: The Mathematics of Black-Hole Mechanics},
  author = {Poisson, Eric},
  publisher = {Cambridge University Press},
  year = {2004},
  doi = {10.1017/CBO9780511606601}
}

@book{FrolovNovikov1998,
  title = {Black Hole Physics: Basic Concepts and New Developments},
  author = {Frolov, Valeri P. and Novikov, Igor D.},
  publisher = {Springer},
  year = {1998},
  doi = {10.1007/978-94-011-5139-9}
}

@article{AlonsoSerrano2018,
  author = {Alonso-Serrano, Ana and Dabrowski, Mariusz P. and Gohar, Hussain},
  title = {Generalized uncertainty principle impact onto the black holes information flux and the sparsity of Hawking radiation},
  journal = {Physical Review D},
  volume = {97},
  number = {4},
  pages = {044029},
  year = {2018},
  doi = {10.1103/PhysRevD.97.044029}
}

@article{Gray2016,
  author = {Gray, Finnian and Schuster, Sebastian and Van-Brunt, Alexander and Visser, Matt},
  title = {The Hawking cascade from a black hole is extremely sparse},
  journal = {Classical and Quantum Gravity},
  volume = {33},
  number = {11},
  pages = {115003},
  year = {2016},
  doi = {10.1088/0264-9381/33/11/115003}
}

\end{document}